\documentclass[twocolumn,pre,floats,aps,amsmath,amssymb]{revtex4}
\usepackage{graphicx}
\usepackage{physics}
\usepackage{tabularx}
\usepackage{amsmath}
\usepackage{float}
\usepackage{hyperref}

\begin{document}
	
	\title{Effects of measures on phase transitions in two cooperative susceptible-infectious-recovered dynamics}
	\author{Adib Khazaee}
	\affiliation{Department of Physics, Sharif University of Technology, Tehran, Iran}
	\author{Fakhteh Ghanbarnejad\footnote{fakhteh.ghanbarnejad@gmail.com}}
	\affiliation{Department of Physics, Sharif University of Technology, Tehran, Iran}
	\affiliation{Chair for Network Dynamics, Institute for Theoretical Physics and Center for Advancing Electronics Dresden (cfaed), Technical University of Dresden, 01062 Dresden, Germany}
	\date{\today}
	
	\begin{abstract}
		In recent studies, it has been shown that a cooperative interaction in a co-infection spread can lead to a discontinuous transition at a decreased threshold. Here, we investigate effects of immunization with a rate proportional to the extent of the infection on phase transitions of a cooperative co-infection. We use the mean-field approximation to illustrate how measures that remove a portion of the susceptible compartment, like vaccination, with high enough rates can change discontinuous transitions in two coupled susceptible-infectious-recovered dynamics into continuous ones while increasing the threshold of transitions. First, we introduce vaccination with a fixed rate into a symmetric spread of two diseases and investigate the numerical results. Second, we set the rate of measures proportional to the size of the infectious compartment and scrutinize the dynamics. We solve the equations numerically and analytically and probe the transitions for a wide range of parameters. We also determine transition points from the analytical solutions. Third, we adopt a heterogeneous mean-field approach to include heterogeneity and asymmetry in the dynamics and see if the results corresponding to homogeneous symmetric case stand.
	\end{abstract}
	
	\maketitle
	
	\section{Introduction}
	\label{sec:intro}
	
	The ongoing pandemic has bolstered arguments for the necessity of devising models for the propagation of infectious diseases which better simulate the process and make more accurate predictions. One of the basic models for simulating an epidemic is the susceptible-infectious-recovered (SIR) model. The SIR model divides the population into three compartments, the susceptible (S), the infectious (I) and the recovered (R), the dynamics of which is governed by the equations $\dot{S}=-\alpha IS$, $\dot{I}=+\alpha IS-\mu I$ and $\dot{R}=\mu I$, neglecting births and natural deaths~\cite{Keeling2008}. $\alpha$ and $\mu$ are the rate of infection and recovery respectively and the size of the compartments usually is scaled in such as way that we have $S+I+R=1$. The model can be generalized to simulate the spread of two diseases. In that case, there are nine compartments whose dynamics is governed by the following equation.
	\begin{equation}
		\frac{dx_i}{dt}=\sum_j({\mu_{ji}x_j-\mu_{ij}x_i})+\sum_{jk}({\alpha_{jik}x_kx_j-\alpha_{ijk}x_ix_k})
		\label{equ:general}
	\end{equation}
	In the above equation, $\mu_{ij}$ is the rate at which the state $i$ recovers to state $j$. Also, $\alpha_{ijk}$ is the rate at which $i$ changes into $j$ due to an infection by  $k$. Compartment sizes are scaled such that $\sum_i{x_i}=1$. Diseases spreading in a population simultaneously can be mutually exclusive, antagonistic or cooperative~\cite{Chen2013}. Cooperative diseases are those that being infected by one of them makes it more probable to get infected by the other. For instance, tuberculosis (TB) cases were increased during the influenza (H1N1) pandemic~\cite{oei2012}. Also, people infected by HIV are more likely to contract hepatitis B and C~\cite{Martcheva2006}, TB~\cite{Sharma2005} or malaria~\cite{AbuRaddad2006}.	
	The subject of simulating the spread of two diseases has interested many researchers through the past two decades. The spread of two competing diseases has been investigated as two sequential bond percolation processes on one contact network or overlay networks to find conditions for the coexistence of the diseases~\cite{Newman2005, Funk2010, Karrer2011}. In another study, a symmetric form of the equations~(\ref{equ:general}) has been proposed to examine the behavior of the dynamics~\cite{Chen2013, Zarei2019}. Solving the mean-field equations shows that discontinuous transitions can happen in the dynamics of a cooperative co-infection. The discontinuous transitions have also been illustrated through simulations on various networks~\cite{Grassberger2016, Cai2015}. It was proved that the symmetric model introduced by Chen et. al. is tantamount to the spatially homogeneous limit of the extended general epidemic process (EGEP)~\cite{Janssen2004} and the corresponding discontinuous transition is a spinodal transition~\cite{Janssen2016}. Some researchers have used the microscopic Markov chain approach to deal with the spread of interacting pathogens or information in multiplex networks~\cite{Arenas2013, Granell2014, Gao2016, Soriano-Panos2019}. Cyclic epidemics of multistrains have been studied~\cite{Zhang2014}. Some works have been done on the mathematical modeling of specific co-infections e.g. malaria-toxoplasmosis~\cite{Ogunmiloro2019, Aldila2018}. Effects of co-infections on pathogen evolution have been examined~\cite{Bushman2019}. Degree-based frameworks have been introduced to approximate the effect of network connectivity structures on the dynamics of co-infections~\cite{Sanz2014, Fennell2019}. Two interacting SIS dynamics have been explored through mean-field approximations and an analytic theory~\cite{Chen2017, Chen2019, Pan2020, Khanjanianpak2020}. The generalized bond percolation has been used to probe the final state of a co-infection dynamic on complex networks with a heterogeneous population~\cite{Zhang2020} and in the presence of self-awareness control~\cite{Wang2019}. In the latter, it has been observed that applying self-awareness control can change discontinuous transitions of a cooperative spread and make them continuous. Some researchers have worked on finding an optimal strategy for vaccination or quarantine for a co-infection through approaches such as the microscopic Markov chain approach~\cite{Ye2021}, the bond percolation process~\cite{Wang2021} and a degree-based mean-field approximation~\cite{Chen2021}.
	
	In the present study, our focus is on effects of immunization with a rate proportional to the extent of the infection on phase transitions of cooperative co-infections.
	As previously mentioned, discontinuous transitions have been observed in the spread of two cooperative diseases through various approaches. By using the bond percolation process, it has also been shown that immunization with a fixed rate can make these discontinuous transitions continuous~\cite{Wang2019}. We confirm this with our approach too.	Here, assuming that the population of the infectious considerably affects the measures policies and people's attitudes toward vaccination, we set the rate at which measures are applied to be proportional to the size of the infectious compartment and scrutinize their effects. We use mean-field approximations to illustrate how measures that remove a portion of the susceptible compartment, like vaccination, with high enough rates can change discontinuous transitions in two coupled SIR dynamics into continuous ones while increasing the threshold of transitions. We mainly investigate the symmetric spread of two diseases in a well-mixed population and extend the exploration to include an asymmetric spread on a heterogeneous population. The structure of this paper is as follows. In section \ref{sec:2SIR}, we review a symmetric form of two coupled SIR dynamics proposed by Chen et al.~\cite{Chen2013}. In section \ref{sec:2SIR_Vaccination}, we include measures into the dynamics by adding a term to the governing equations which reduces the size of the susceptible at a certain fixed rate. Moreover, we explore the resulting dynamics through numerical experimentation. In section \ref{sec:analytical}, assuming that the size of the infectious compartment dictates the rate at which the measures are applied, we set a variable rate for measures proportional to the population of the infectious and look into the results of numerical experimentation. Also, we solve the governing equations analytically and scrutinize the characteristics of the dynamics through analytical solutions. In section \ref{sec:DBG}, we include heterogeneity in the dynamics (using a heterogeneous mean-field approximation) to make sure that the observed effects of measures are not merely limited to a mixed population or symmetric dynamics. In the last section, we sum up the results and discussions.
	
	\section{Symmetric framework of two coupled susceptible-infectious-recovered dynamics}
	\label{sec:2SIR}

	In a previous study~\cite{Chen2013}, a symmetric framework was proposed to simplify governing equations~(\ref{equ:general}) in order to investigate the characteristics of the dynamics. In the proposed framework, the population is divided into nine groups denoted $[S], [A], [a], [B], [b], [AB], [Ab], [aB]$ and $[ab]$, the sum of which is equal to $1$. $[S]$ represents the susceptible. Uppercase letters correspond to the portion of the population infected by diseases $A$ or $B$ while lowercase letters correspond to the portion recovered from the diseases. Primary and secondary infection rates of both diseases are assumed equal. Moreover, the recovery rates are all set to 1 (figure~\ref{fig:SymmDiag}). Using the symmetry, Chen et al. defined three variables, $S$, $P$, and $X$ to summarize the nine governing equations and ended up with the following equations~\cite{Chen2013}:
	
	\begin{figure}[htbp]
		\begin{center}
			\includegraphics[width=3.5in]{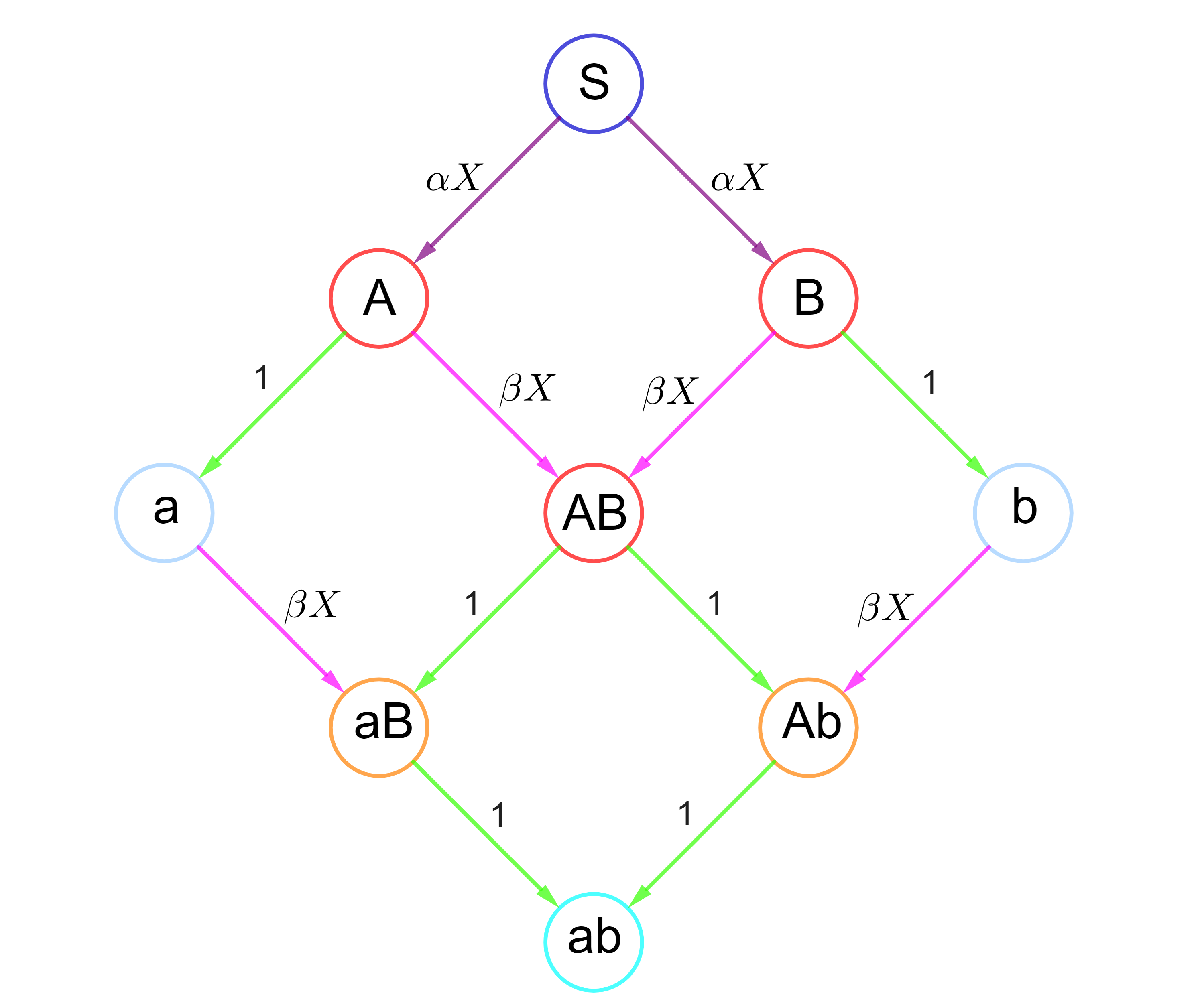}
			\caption{A schematic diagram of two coupled SIR dynamics with symmetric parameters and initial conditions. $S$: the susceptible. $A$, $B$, and $AB$: infected by one or two diseases. $Ab$: infected by $A$ and recovered from $B$. $aB$: Infected by $B$ and recovered from $A$. $a$, $b$, and $ab$: Recovered from one or two diseases. $X$ is defined as $X=[A]+[AB]+[Ab]=[B]+[AB]+[aB]$. Purple lines: getting infected for the first time with the rate $\alpha$. Pink lines: getting infected by the second disease with the rate $\beta$. Green lines: being recovered with the rate 1.}
			\label{fig:SymmDiag}
		\end{center}
	\end{figure}

	\begin{figure*}[htbp]
		\begin{center}
			\includegraphics[width=6.5in]{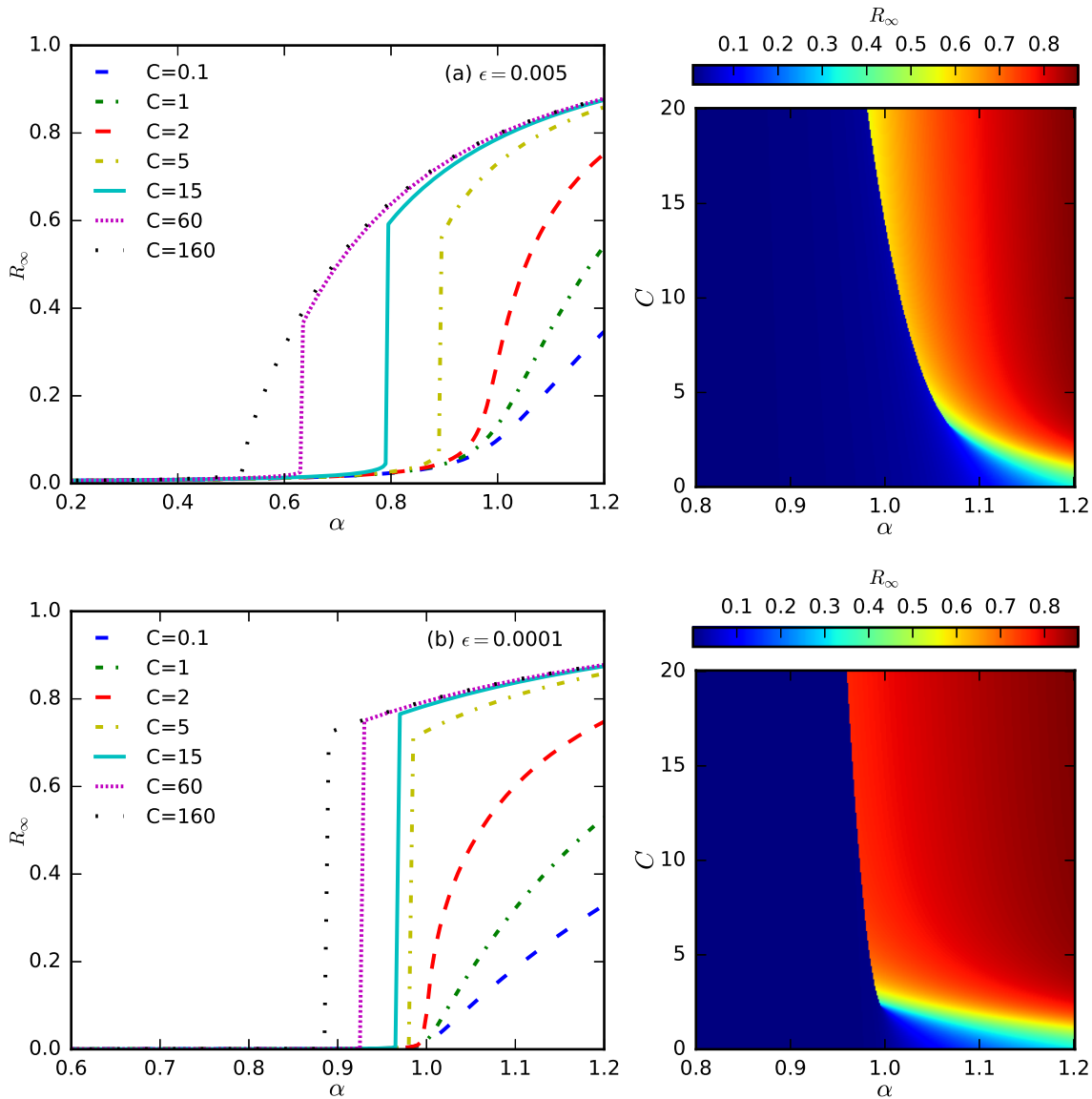}
			\caption{Phase transitions in the symmetric case of two coupled SIR dynamics~\cite{Chen2013}. The final size of the recovered, $R_\infty=1-S_\infty$, is depicted against the rate of the first infection $\alpha$ for different levels of cooperativity $C=\frac{\beta}{\alpha}$. In the left panel, diagrams are illustrated for some specific values of $C$, while the heat map of $R_\infty$ for a range of $C$ and $\alpha$ is shown on the right panel. The model parameters and initial conditions are $\beta=C\alpha$, $X_0=P_0=\frac{\epsilon}{2}$ and $S_0=1-\epsilon$ (equations~(\ref{eq:SPX})).}
			\label{fig:2SIR}
		\end{center}
	\end{figure*}

	\begin{equation}
		\begin{array}{l} \dot{S}=-2\alpha SX, \\ 
			\dot{P}=(\alpha S-\beta P)X, \\ 
			\dot{X}=(\alpha S+\beta P-1)X, 
		\end{array}
		\label{eq:SPX}
	\end{equation}
	in which we have:
	\begin{equation}
		\begin{array}{l} S=[S], \\ 
			P=[A]+[a]=[B]+[b], \\ 
			X=[A]+[AB]+[Ab]=[B]+[AB]+[aB].
		\end{array}
	\end{equation}

	The parameters $\alpha$ and $\beta$ are the rate of primary and secondary infections respectively. Numerical solutions to the equations (\ref{eq:SPX}) are depicted for two different initial levels of infections in figure~\ref{fig:2SIR} where the parameter $C$, defined as $C=\frac{\beta}{\alpha}$, is the representation of the cooperativity. The results show that when $\beta>\alpha$ the transition occurs for $\alpha$ less than 1. Furthermore, when $\beta>2\alpha$, discontinuous transitions are observed for $0.5<\alpha<1$~\cite{Chen2013}. Incorporating control measures evidently affects the dynamics. In the following sections, we show how certain control measures such as vaccination can turn these discontinuous transitions into continuous transitions.

	\section{Measures with a constant rate in the symmetric framework of two coupled susceptible-infectious-recovered dynamics}
	\label{sec:2SIR_Vaccination}
	
\begin{figure}[htbp]
		\begin{center}
			\includegraphics[width=3.5in]{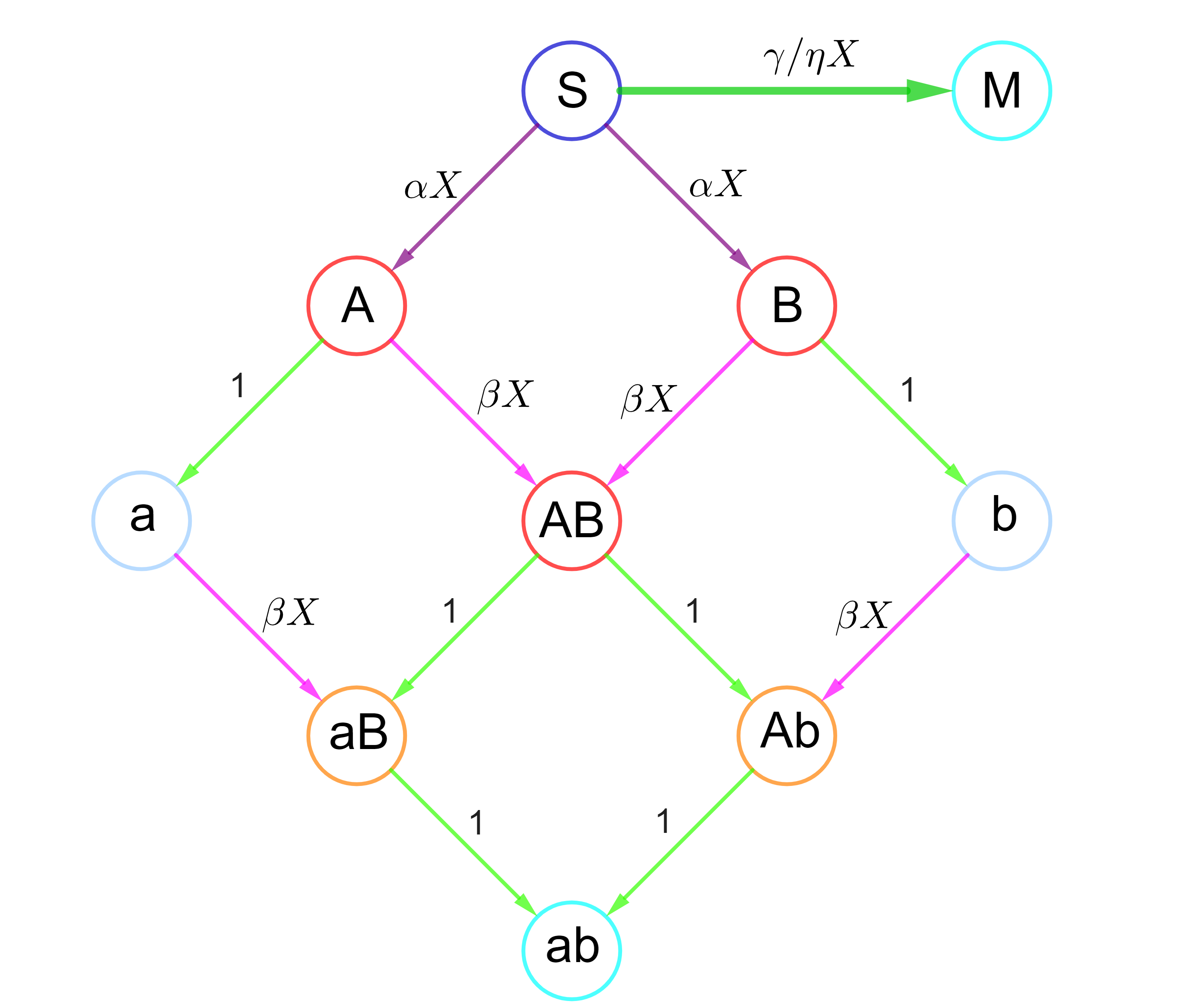}
			\caption{A schematic diagram of introducing control measures into two coupled SIR dynamics with symmetric parameters and initial conditions. $M$: the vaccinated. $\gamma$ or $\eta X$: the rate of measures. Other compartments and parameters are the same as in figure~\ref{fig:SymmDiag}.}
			\label{fig:SymmDiagV}
		\end{center}
	\end{figure}

	Here, we introduce measures into two coupled SIR dynamics and investigate how this can change the transitions. We assume that the susceptible getting vaccinated are being removed from the compartment $[S]$ at the rate $\gamma$ and added directly to the compartment $[M]$ which denotes the vaccinated portion of the population, as shown in figure~\ref{fig:SymmDiagV}. Thus, we have $[S]+[A]+[a]+[B]+[b]+[AB]+[Ab]+ [aB]+[ab]+[M]=1$. This way, the symmetry is preserved and we can include measures in the governing equations as follows ($M=[M]$).
	
	\begin{equation}
		\begin{array}{l} \dot{S}=-2\alpha SX-\gamma S, \\ 
			\dot{P}=(\alpha S-\beta P)X, \\ 
			\dot{X}=(\alpha S+\beta P-1)X, \\ 
			\dot{M}=\gamma S. 
		\end{array}
		\label{eq:SPXV}
	\end{equation}
	
	\begin{figure*}[htbp]
		\begin{center}
			\includegraphics[width=6.5in]{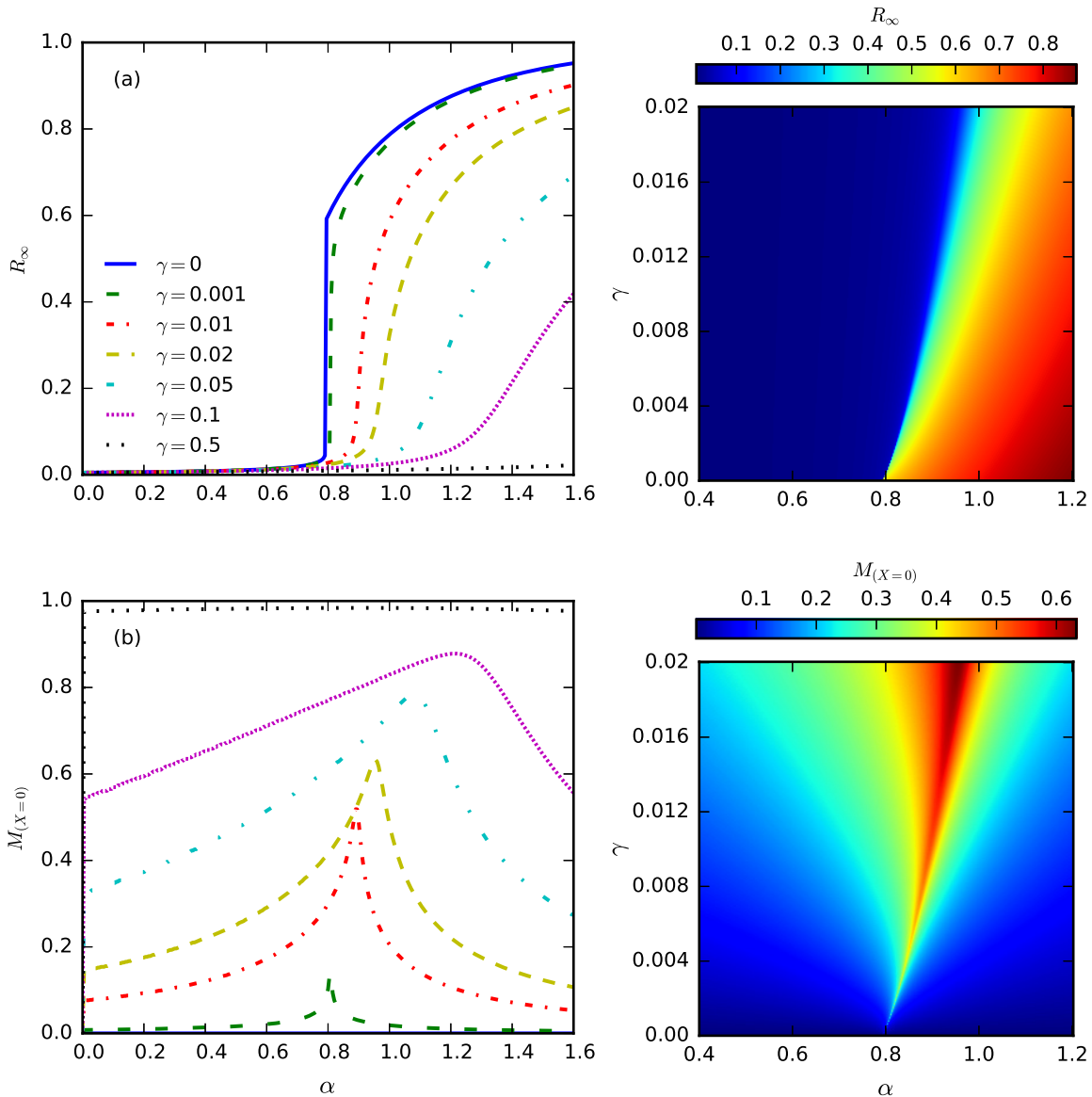}
			\caption{The effect of measures with a fixed rate on the phase transitions in the symmetric case of two coupled SIR dynamics. The final size (at $X=0$) of the recovered compartment, (a) $R_\infty=1-S_{(X=0)}-M_{(X=0)}$, and the vaccinated compartment, (b) $M_{(X=0)}$, are depicted against the rate of the first infection $\alpha$ for different measure rates $\gamma$. The model parameters and initial conditions are $\epsilon=0.005$, $\beta=15\alpha$, $X_0=P_0=\frac{\epsilon}{2}$, and $S_0=1-\epsilon$ (equations~(\ref{eq:SPXV})). On the left panel, diagrams are illustrated for some specific values of $\gamma$, while heat maps of $R_\infty$ and $M_{(X=0)}$ are shown on the right panel, in which the range of $\gamma$ is limited to the region of the change in the transition type.}
			\label{fig:RVvsAlpha}
		\end{center}
	\end{figure*}

\begin{figure}[htbp]
	\begin{center}
		\includegraphics[width=3.7in]{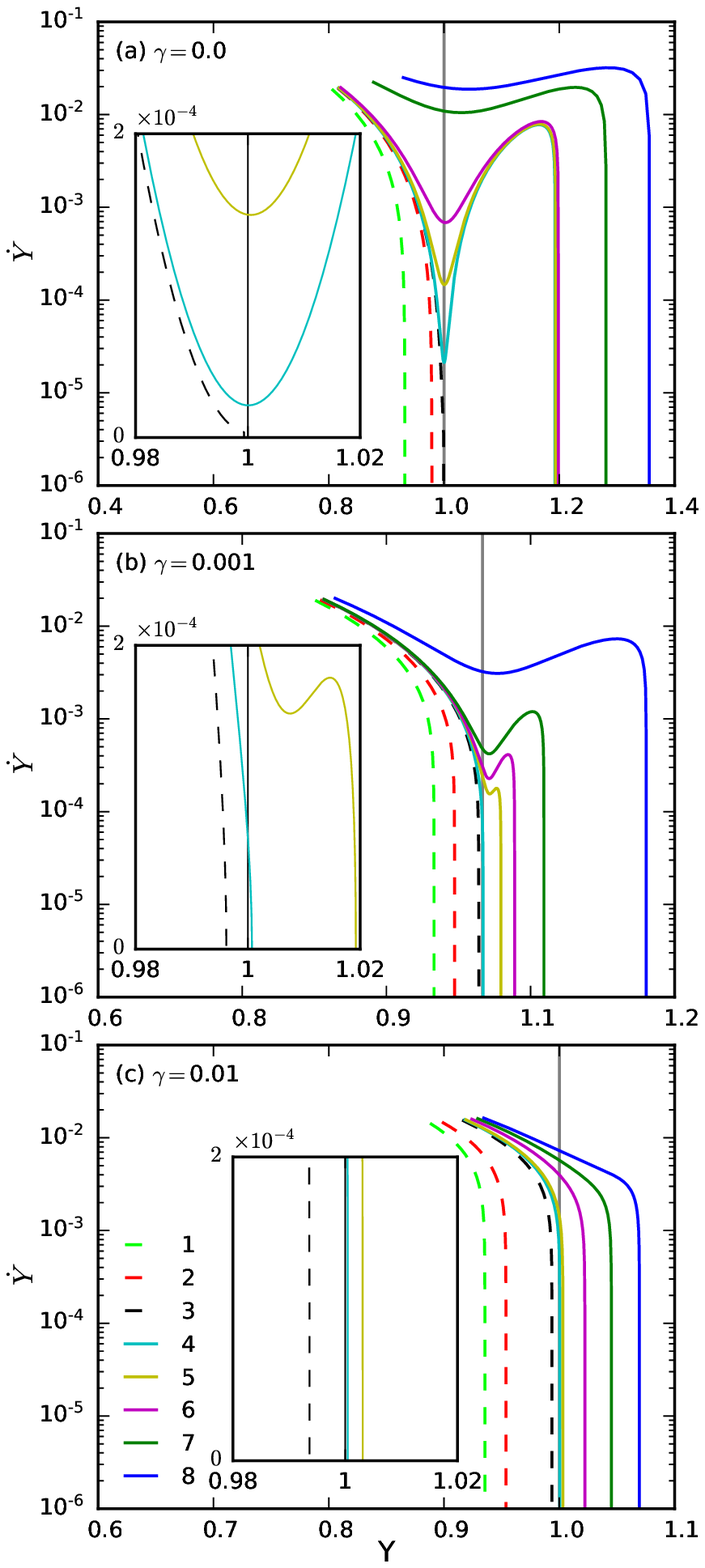}
		\caption{Near-threshold trajectories of $\dot{Y}(t)$ are depicted against $Y(t)$ for different values of $\gamma$. The model parameters and initial conditions are $\beta=15\alpha$, $X_0=P_0=\frac{\epsilon}{2}$, $S_0=1-\epsilon$ and $\epsilon=0.005$ (equations~(\ref{eq:SPXV})). Numbers 1-8 in the legend correspond to values of $\alpha$ as follows. (a) $\alpha \in \{0.78, 0.79, 0.7912, 0.7913, 0.792, 0.795, 0.85, 0.90\}$, (b) $\alpha \in \{0.80, 0.805, 0.8081, 0.8083, 0.8085, 0.8086, 0.809, 0.82\}$, (c) $\alpha \in \{0.86, 0.87, 0.887, 0.8892, 0.89, 0.895, 0.90, 0.905\}$ . Dashed and solid lines correspond to cases with $\alpha$ below and above the threshold respectively.}
		\label{fig:YdotvsY}
	\end{center}
\end{figure}

	We solved the above equations numerically and explored how vaccination affects transitions. The results corresponding to the case in which $\beta=15\alpha$ (the solid cyan curve in figure~\ref{fig:2SIR}) are illustrated in figure~\ref{fig:RVvsAlpha}. The initial conditions are $X_0=P_0=\frac{\epsilon}{2}$, and $S_0=1-\epsilon$ and $\epsilon=0.005$. In the numerical experiments, the dynamic was stopped when the infectious compartment diminishes ($X=0$) and the sizes of immune compartments (recovered from the infections or vaccinated) have been recorded for different values of $\alpha$ and $\gamma$. Note that since the measure rate is constant, if the dynamic is not stopped, it can go on until the remainder of the susceptible is vaccinated. We call the recorded size of the measure compartment $M_{(X=0)}$ to emphasize this fact. Figuratively, there is a competition between the diseases and control measures to immunize the population while the infectious compartments are active. 
	
	As indicated in the results, not only do measures increase the threshold, but they also turn discontinuous transitions into continuous ones with high enough rates ($\gamma$). For $\epsilon=0.005$, when $\gamma$ reaches 0.01, the threshold goes from around 0.79 to around 0.9 and the transition becomes continuous (figure~\ref{fig:RVvsAlpha}a). Moreover, the maximum of $M_{(X=0)}$ occurs at the transition threshold, as shown in figure~\ref{fig:RVvsAlpha}b. The dynamic has similar behavior for relatively small initial infections, $\epsilon=0.0001$. We have included the results of such a case in the appendix (figure~\ref{fig:RVvsAlpha1}). There, the transition gets continuous when $\gamma=0.005$ for which the threshold has reached 1.1 from around 1 in the absence of measures.	
	To distinguish between disease free and epidemic phases, and also continuous and discontinuous transitions, we follow the argument made by Chen et al.~\cite{Chen2013}. If we define $Y=\alpha S+\beta P$, we will have $\dot{X}=(Y-1)X$ which means that the dynamic of $X$ depends directly on whether $Y$ stays below $1$ or surpasses it. The rate of $Y$ can be rewritten as follows.
	\begin{equation}
		\begin{array}{l}
			\dot{Y}=\left[(\beta-2\alpha)\alpha S -\beta^2 P\right]X-\gamma\alpha S \\
			=\left[2(\beta-\alpha)\alpha S -\beta Y\right]X-\gamma\alpha S.
		\end{array}
		\label{equ:Ydot}
	\end{equation}	  
	For $\beta<2\alpha$, $\dot{y}$ is negative and the outbreak occurs only for values of $\alpha$ which are greater than 1 making $Y(0)>1$. However, we are interested in transitions that happen as a result of the cooperation of diseases ($\beta\gg\alpha$). For $\alpha<1$ and $\epsilon\ll1$, $Y$ starts from a value less than 1, therefore, an outbreak occurs only for cases in which $Y$ reaches 1 before the infection is dead ($X=0$). As argued by Chen et al.~\cite{Chen2013}, for $\alpha$ slightly above the threshold, when $Y$ reaches 1, if $\dot{Y}=0$, the transition is continuous. Otherwise, it is discontinuous. Equation~(\ref{equ:Ydot}) shows that measures directly affect $\dot{Y}$, which explains how they can prevent the outbreak or change the type of the transition. This reasoning can be illustrated by following how the path which the dynamic goes through in the $Y-\dot{Y}$ plane behaves when it reaches the line $Y=1$ at the vicinity of the threshold. The near-threshold trajectories of $(Y,\dot{Y})$ are depicted in figure~\ref{fig:YdotvsY}. The trajectories illustrate how increasing $\gamma$ can change the intersection of the $\dot{Y}(Y)$ curve and the $Y=1$ line. The dashed and solid lines pertain to cases with $\alpha$ below and above the threshold respectively. As shown in figures~\ref{fig:YdotvsY}a and \ref{fig:YdotvsY}b, the trajectories corresponding to the transition $\alpha$ cross the line $Y=1$ at $\dot{Y}>0$. On the other hand, for $\gamma$=0.01 in figure~\ref{fig:YdotvsY}c, the transition trajectory becomes tangent to $Y=1$ and goes to zero, corresponding to a continuous transition. We have also looked at near-threshold trajectories of $(Y,X)$ in figure~\ref{fig:XvsY} presented in the appendix. We can see that the trajectories corresponding to the transition $\alpha$ goes considerably over the line of $Y=1$ for discontinuous transitions (figures~\ref{fig:XvsY}a and \ref{fig:XvsY}b). Meanwhile, for a continuous transition, the transition trajectory hits the line of $Y=1$ and comes back (figure~\ref{fig:XvsY}a).

	\section{Measures with a rate proportional to the infectious population in the symmetric framework of two coupled susceptible-infectious-recovered dynamics}
	\label{sec:analytical}
	
	 In the middle of a catastrophe such as a pandemic, there are usually a variety of issues which need to be addressed. These issues would not necessarily get prioritized at the same time. Sometimes, the issue that emerges as an immediate threat gets the most of the attention for a while and then gets behind other priorities after it is relatively mitigated. Therefore, the control measures might be applied at a variable rate which depends on the seriousness of the spread represented by the population of the infectious. Moreover, when we talk about vaccination, the will of the people also comes into the picture. The existence of more infectious people could encourage others to get vaccinated. In these cases, the size of the infectious compartment can dictate the rate at which the control measures are applied. To investigate such a dynamic, we can set the measure rate proportional to the infectious compartment ($\gamma=\eta X$, figure~\ref{fig:SymmDiagV}) and rewrite the governing equations as follows.
	
	\begin{equation}
		\begin{array}{l} \dot{S}=-2\alpha SX-\eta XS, \\
			\dot{P}=(\alpha S-\beta P)X, \\
			\dot{X}=(\alpha S+\beta P-1)X, \\
			\dot{M}=\eta XS.
		\end{array}
		\label{eq:SPXVEtaX}
	\end{equation}	
	
\begin{figure*}[htbp]
	\begin{center}
		\includegraphics[width=6.5in]{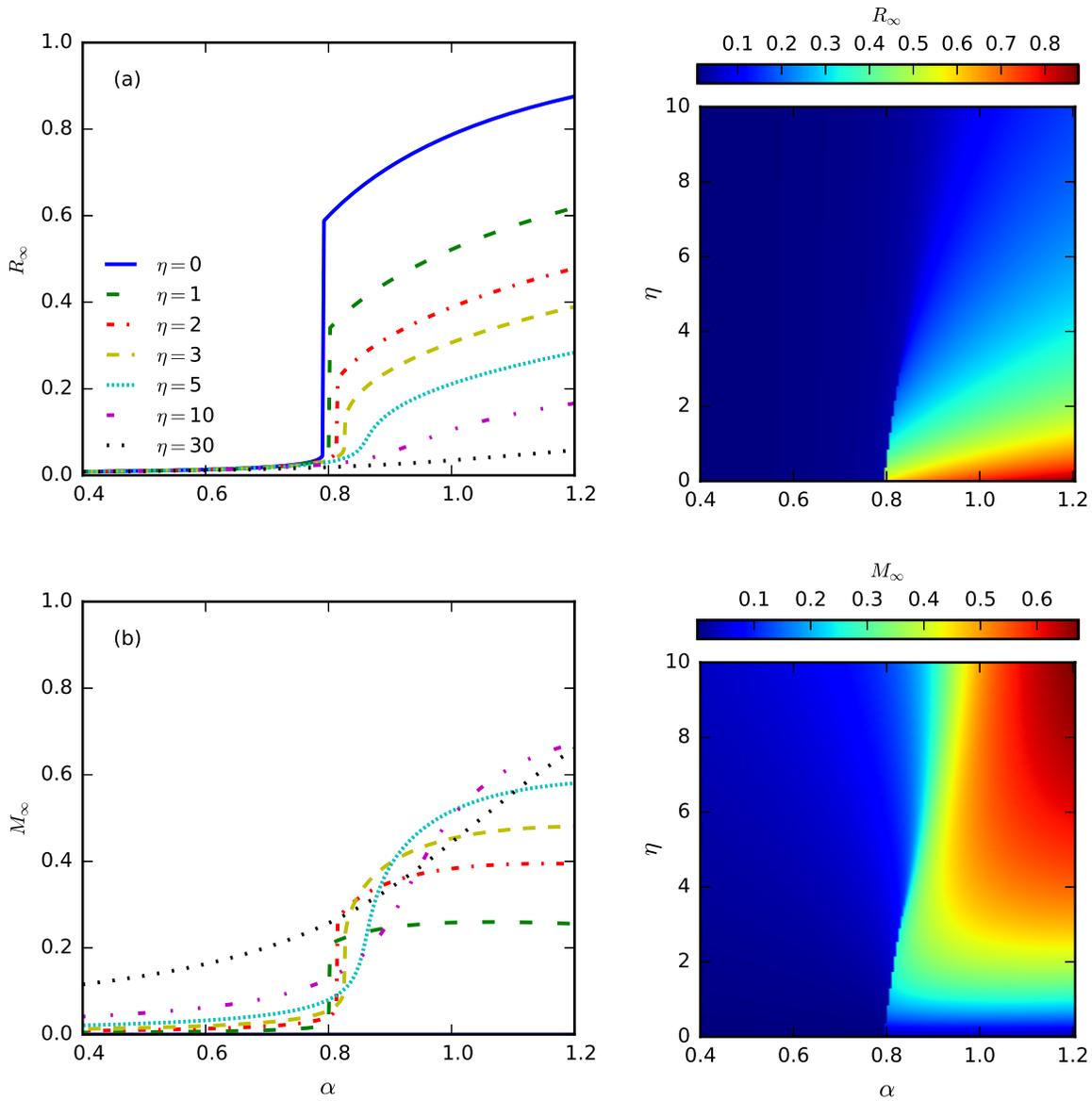}
		\caption{The effect of measures with a rate proportional to the size of the infectious compartment on the phase transitions in the symmetric case of two coupled SIR dynamics. The final size (at the end of the dynamics) of the recovered compartment, (a) $R_\infty=1-S_\infty-M_\infty$, and the vaccinated compartment, (b) $M_\infty$, are depicted against the rate of the first infection $\alpha$ for different measure rates $\eta$. The model parameters and initial conditions are $\epsilon=0.005$, $\beta=15\alpha$, $X_0=P_0=\frac{\epsilon}{2}$, and $S_0=1-\epsilon$ [equations~(\ref{eq:SPXVEtaX})]. On the left panel, diagrams are illustrated for some specific values of $\eta$, while heat maps of $R_\infty$ and $M_\infty$ are shown on the right panel, in which the range of $\eta$ is limited to the region of the change in the transition type.}
		\label{fig:RVvsAlphaEtaX}
	\end{center}
\end{figure*}

	The above equations have been solved numerically with the same initial conditions as the previous section, $X_0=P_0=\frac{\epsilon}{2}$, $S_0=1-\epsilon$, and $\epsilon=0.005$. The results, depicted in figure~\ref{fig:RVvsAlphaEtaX}, illustrate that effects of measures on transitions are relatively similar to those of fixed-rate measures. With a high enough rate [between $\eta=3$ and $\eta=5$ in figure~\ref{fig:RVvsAlphaEtaX}(a)], measures can turn a discontinuous transition into a continuous one. In comparison with the fixed-rate measures,  however, the increase in the threshold due to the measures is lower. With fixed-rate measures [figure~\ref{fig:RVvsAlpha}(a)], when the measure rate ($\gamma$) gets so high that the transitions becomes continuous, the threshold reaches around 0.9 (from 0.79 for no measure). Meanwhile, with variable-rate measures [figure~\ref{fig:RVvsAlphaEtaX}(a)], when the measure rate ($\eta$) reaches a level which can turn the discontinuous transition into a continuous one, the threshold is 0.83.
Another difference between effects of fixed-rate measures and variable-rate measures shows up in the size of the measure compartment. With fixed-rate measures [figure~\ref{fig:RVvsAlpha}(b)] where $\dot{M}=\gamma S$, since the rate is fixed and we stop measures at the moment when $X=0$, the size of measure compartment peaks at the vicinity of the threshold. Because, for lower $\alpha$, there is less time for vaccination and for higher $\alpha$, the disease acts faster in depleting the susceptible. On the other hand, with variable-rate measures [figure~\ref{fig:RVvsAlphaEtaX}(a)] where $\dot{M}=\eta XS$, $M_{\infty}$ reaches a plateau for $\alpha$ higher that the threshold, which is a representative of the fact that the rate of measures is proportional to $X$. We have also solved the governing equations numerically for the initial conditions $X_0=P_0=\frac{\epsilon}{2}$ and $S_0=1-\epsilon$ with a different $\epsilon=0.0001$, depicted in figure~\ref{fig:RVvsAlphaEtaX1} of the appendix. The characteristics of the results are fairly similar. The transition becomes continuous for a measure rate between $\eta=10$ and $\eta=12.5$, while the threshold increases to 0.986 from 0.965 without measures.

\begin{figure*}[htbp]
	\begin{center}
		\includegraphics[width=6.5in]{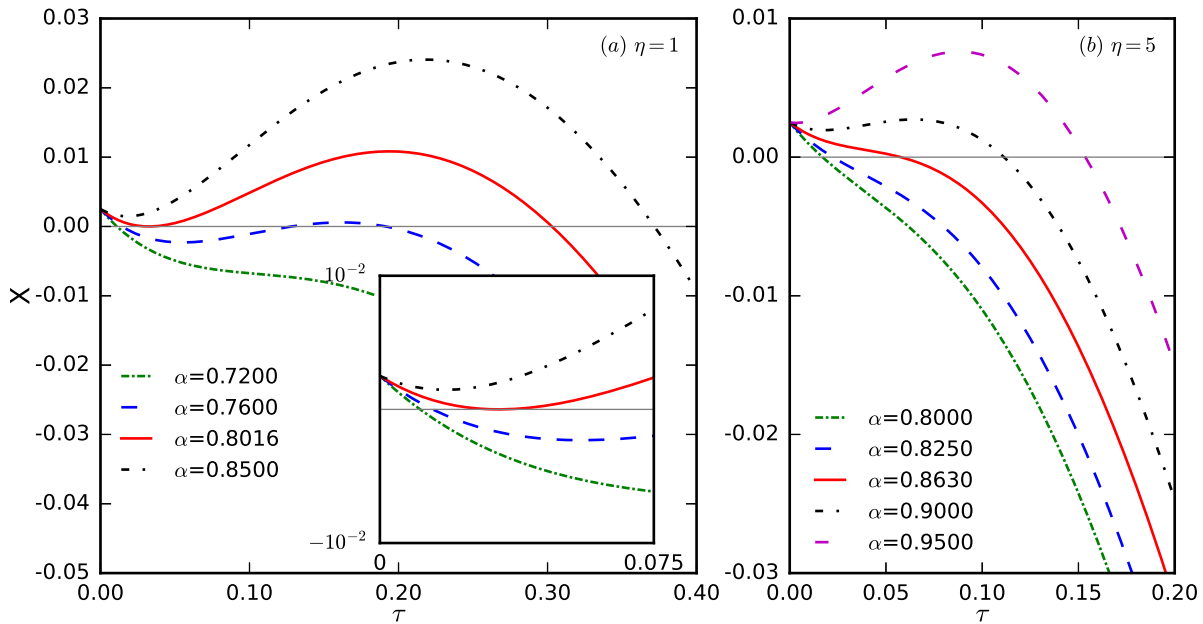}
		\caption{The phase portrait of $\tau$ as an order parameter against $\dot{\tau}=X$ [equation~(\ref{eq:X})]. The diagram illustrates changes of fixed points of $\tau$ around the transition $\alpha$ (solid red line) for two different types of transitions, (a) a discontinuous transition corresponding to a saddle-node-like bifurcation, and (b) a continuous transition. $\beta=15\alpha$, $X_0=P_0=\frac{\epsilon}{2}$, $S_0=1-\epsilon$, and $\epsilon=0.005$.}
		\label{fig:XTauBifur}
	\end{center}
\end{figure*}

\begin{figure}[htbp]
	\begin{center}
		\includegraphics[width=3.5in]{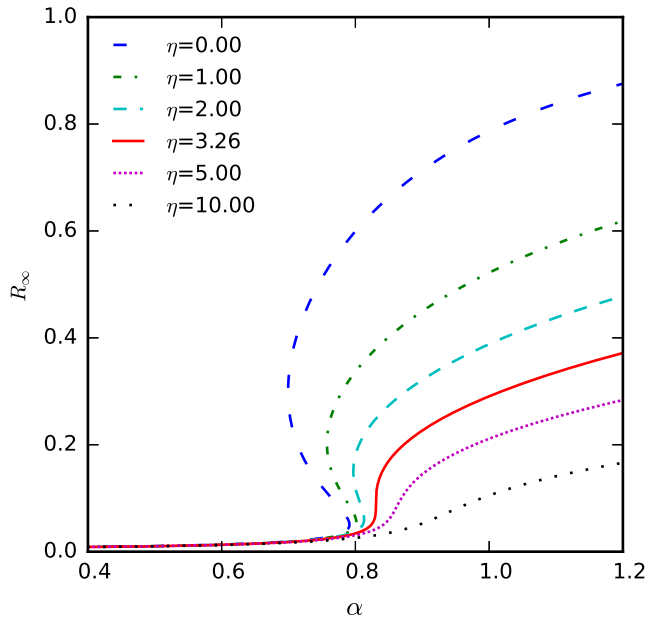}
		\caption{Illustration of how increasing $\eta$ diminishes the part of $R_\infty(\alpha)$ with two stable fixed points. The final size of the recovered compartment, $R_\infty=1-S_\infty-M_\infty$, determined through analytical equations~(\ref{eq:AnalSV}),~(\ref{eq:AnalP}) and~(\ref{eq:X}), is depicted against the rate of the first infection $\alpha$ for different measure rates $\eta$. The model parameters are $\beta=15\alpha$, $X_0=P_0=\frac{\epsilon}{2}$, $S_0=1-\epsilon$ and $\epsilon=0.005$. The solid red curve corresponds to $\eta^*$, the threshold of $\eta$ at which the transition gets continuous.}
		\label{fig:AnalRvsAlpha}
	\end{center}
\end{figure}

\begin{figure}[htbp]
	\begin{center}
		\includegraphics[width=3.5in]{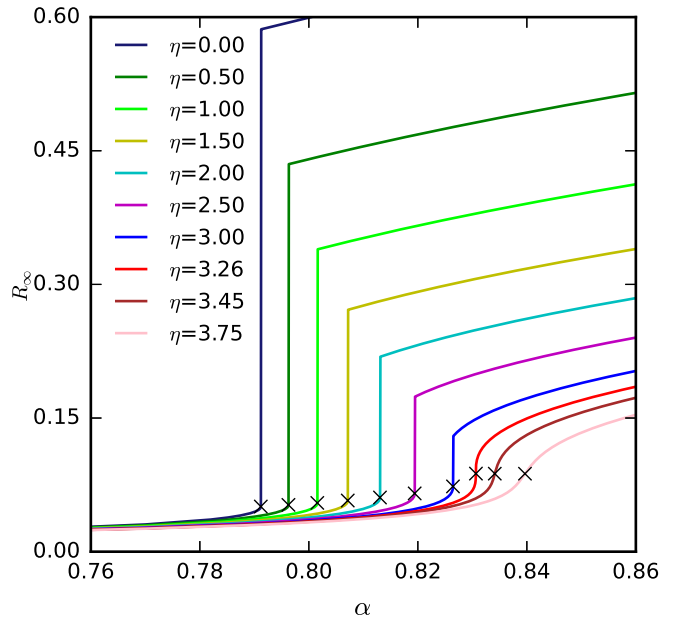}
		\caption{Illustration of transition points determined through analytical solution of the governing equations on the diagram of phase transitions. The final size (at the end of the dynamic) of the recovered compartment, $R_\infty=1-S_\infty-M_\infty$, is depicted against the rate of the first infection $\alpha$ for different measure rates $\eta$. Transition points are depicted by $\times$ markers. The model parameters are $\beta=15\alpha$, $X_0=P_0=\frac{\epsilon}{2}$, $S_0=1-\epsilon$, and $\epsilon=0.005$. The solid red curve corresponds to $\eta^*$, the threshold of $\eta$ at which the transition gets continuous.}
		\label{fig:Tricirtical}
	\end{center}
\end{figure}
	
	Equations~(\ref{eq:SPXVEtaX}) can also be solved analytically. Using the approach proposed by previous studies ~\cite{Zarei2019, Janssen2016}, we can define $d\tau=Xdt$ and turn the equations~(\ref{eq:SPXVEtaX}) into the following linear form:
	
	\begin{equation}
		\frac{dS}{d\tau}=-2\alpha S-\eta S
		\label{eq:STau}	
	\end{equation}	
	\begin{equation}
		\frac{dP}{d\tau}=\alpha S-\beta P	
		\label{eq:PTau}	
	\end{equation}		
	\begin{equation}
		\frac{dX}{d\tau}=\alpha S+\beta P-1	
		\label{eq:XTau}	
	\end{equation}	
	\begin{equation}
		\frac{dM}{d\tau}=\eta S	
		\label{eq:MTau}	
	\end{equation}	

	We can solve these equations to determine $S$, $P$, $X$, and $M$ as functions of $\tau$. Integrating the equation~(\ref{eq:STau}) leads to $S$. Inserting $S$ into the equation~(\ref{eq:PTau}) and solving the resulting differential equation give us $P$. We can find $M$ by inserting S in the equation~(\ref{eq:MTau}) and integrating the result. Following the above steps culminates in the following relations.
	
	\begin{equation}
		\def\arraystretch{1.5}
		\begin{array}{l} 	
			S=S_0\exp[(-2\alpha-\eta)\tau], \\
			M=\frac{\eta S_0}{2\alpha+\eta}\biggl(1-\exp[(-2\alpha-\eta)\tau]\biggr), \\
		\end{array} 
		\label{eq:AnalSV}
	\end{equation}	
	
	\begin{multline}
		P=P_0\exp[-\beta\tau] \\
		-\frac{\alpha S_0}{\beta-(2\alpha+\eta)}\biggl(\exp[-\beta\tau]-\exp[(-2\alpha-\eta)\tau]\biggr).
		\label{eq:AnalP}
	\end{multline}	

	To determine $X$, we introduce a new variable $U=[a]+[aB]+[ab]=[b]+[Ab]+[ab]$ (inspired by~\cite{Zarei2019}). It is easy to see that $S+P+X+U+M=1$. On the other hand, from equations~(\ref{eq:SPXVEtaX}), we can see that $\dot{U}=-(\dot{S}+\dot{P}+\dot{X})-\dot{M}=X$ which leads to $U=\tau$. Therefore, we have $S+P+X+\tau+M=1$, which evidently results in the following relation for $X$. 
	
	\begin{equation}
		X=1-\tau-M-P-S
		\label{eq:X}
	\end{equation}		
	
	Zarei et al.~\cite{Zarei2019} argued that the discontinuous transitions are the result of a saddle-node-like bifurcation. The same argument is valid here. Using the above relations, one can see that $\tau=[a]+[aB]+[ab]$, which means that when the infection is dead ($X=0$), we have $\tau_\infty=[a]_\infty+[ab]_\infty$. On the other hand, the recovered compartment is $R_\infty=2[a]_\infty+[ab]_\infty$. Hence, we can also look at $\tau_\infty$, instead of $R_\infty$, as the order parameter and explore the transitions. Furthermore, $\dot{\tau}=X$ is resulted from the definition of $\tau$ and we have determined $X(\tau)$. Although we do not have an analytical solution to the equation $X(\tau)=0$, we can scrutinize what happens to the fixed points of $\tau$ around the transition point through plotting $X$ against $\tau$ for different values of the control parameter $\alpha$. As shown in figure~\ref{fig:XTauBifur}(a) representing a discontinuous transition, the equation $X(\tau)=0$ has only one root for values of $\alpha$ less than the threshold (the solid red curve) and relatively far from it. When we increase $\alpha$, as we get close to the threshold, other fixed points show up. The first fixed point is of the order of $\epsilon$ while the last one goes towards the order of 1 as $\alpha$ increases. In these cases, although there are more than one fixed point, the dynamic always ends up in the one closest to the origin. At the threshold, the curve becomes tangent to the line of $X=0$ at the first fixed point, which means that $X=0$ and $X_{\tau}=0$ occur simultaneously. When $\alpha$ passes the threshold, the first fixed point disappears and $\tau$ jumps from the first fixed point to the second, which emerges as a discontinuous transition. In the context of dynamical systems, bifurcations in which one fixed point turns into two fixed points and vice versa are called saddle nodes. Here, only one of the two emerged fixed points is physical, which makes the bifurcation a saddle-node-like one. In figure~\ref{fig:XTauBifur}(b), on the other hand, increasing $\alpha$ never results in more than one fixed point, which shows how measures with a high-enough rate can eliminate the bifurcation and make the transition continuous. In figure~\ref{fig:XTauBifur}, $\epsilon$ in the initial conditions has been set to 0.005. We have shown the bifurcation for the case of $\epsilon=0.0001$ in figures~\ref{fig:XTauBifur1}(a) and figures~\ref{fig:XTauBifur1}(b) of the appendix.

	We can also determine $R_\infty$ as a function of $\alpha$ by using analytical equations~(\ref{eq:AnalSV}),~(\ref{eq:AnalP}), and~(\ref{eq:X}). As mentioned before, we have $R_\infty=\tau_\infty+[a]_\infty$. Moreover, at the end of the dynamic, we can write $P_\infty=[a]_\infty$ which means that $R_\infty=\tau_\infty+P_\infty$. We can find $\tau_\infty$ by putting $X=0$ into equation~(\ref{eq:X}) and solving the resulting equation numerically. Inserting the calculated $\tau_\infty$ into equation~(\ref{eq:AnalP}), we find $P_\infty$ culminating in the determination of $R_\infty$ which is depicted against $\alpha$ in figure~\ref{fig:AnalRvsAlpha} [and figure ~\ref{fig:XTauBifur1}(c) of the appendix] for different values of $\eta$. The solid red curves correspond to measure rates at which the transition gets continuous. $\eta^*$ denotes such rates henceforth. As shown figure~\ref{fig:AnalRvsAlpha}, for $\eta<\eta^*$, diagrams of $R_\infty(\alpha)$ have a part in which there are three values of $R_\infty(\alpha)$ for each $\alpha$ less than $\alpha_{tr}$, the transition $\alpha$ corresponding to the onset of transitions. Since only positive-gradient branches of the diagrams are stable~\cite{Janssen2016}, there are two stable fixed points in those parts. For these values of $\alpha$, however, the dynamic always ends up in the lower branch making values of $R_\infty(\alpha)$ over the lower branch not physical. At $\alpha_{tr}$, the lower branch suddenly disappears and a jump occurs in $R_\infty$, representing a discontinuous transition. Evidently, increasing $\eta$ diminishes the part of $R_\infty(\alpha)$ curves with two stable fixed points. When $\eta$ reaches $\eta^*$, that part disappears completely which means that the transition becomes continuous.
	
	It is also interesting to find $\alpha_{tr}$ and $\eta^*$. Setting $X=0$ in the equation~(\ref{eq:X}), we end up with an implicit equation, $F(\tau, \alpha, \eta)=1-\tau-M(\tau, \alpha, \eta)-P(\tau, \alpha, \eta)-S(\tau, \alpha, \eta)=0$, for $\tau_\infty$ as a function of $\alpha$ and $\eta$. For $\eta<\eta^*$, the solution to simultaneous equations $F=0$ and $F_\tau=0$ gives us $\alpha_{tr}$ ($\times$ markers in figure~\ref{fig:Tricirtical}). These equations are equivalent to the simultaneity of $X=0$ and $X_{\tau}=0$, which is in agreement with the saddle-node-like bifurcation explained above. For $\eta>\eta^*$, the inflection point of the $R_\infty(\alpha)$ curve is determined by solving simultaneous equations $F=0$ and $F_\alpha(2F_{\alpha\tau}F_\tau-F_\alpha F_{\tau\tau})-F_\tau^2F_{\alpha\alpha}=0$. The inflection points have been calculated for different values of $\eta$ and are depicted as $\times$ markers on diagrams of phase transitions in figure~\ref{fig:Tricirtical}. At $\eta=\eta^*$, $R_\infty(\alpha)$ is tangent to a vertical line at the inflection point. Therefore, $\eta^*$ comes from the solution to three simultaneous equations $F=0$, $F_\tau=0$ and $F_{\tau\tau}=0$. Solving these equations numerically, we have found $\eta^*$ for $\epsilon=0.005$ in the initial conditions, the solid red curve in figure~\ref{fig:Tricirtical}. The thresholds $\alpha_{tr}$ and $\eta^*$ have been also determined for a lower level of initial infections, $\epsilon=0.0001$, depicted in figure~\ref{fig:XTauBifur1}(d) of the appendix.
	
	\section{Effects of measures in a heterogeneous population}
	\label{sec:DBG}
	
	\begin{figure}[b]
		\begin{center}
			\includegraphics[width=3.5in]{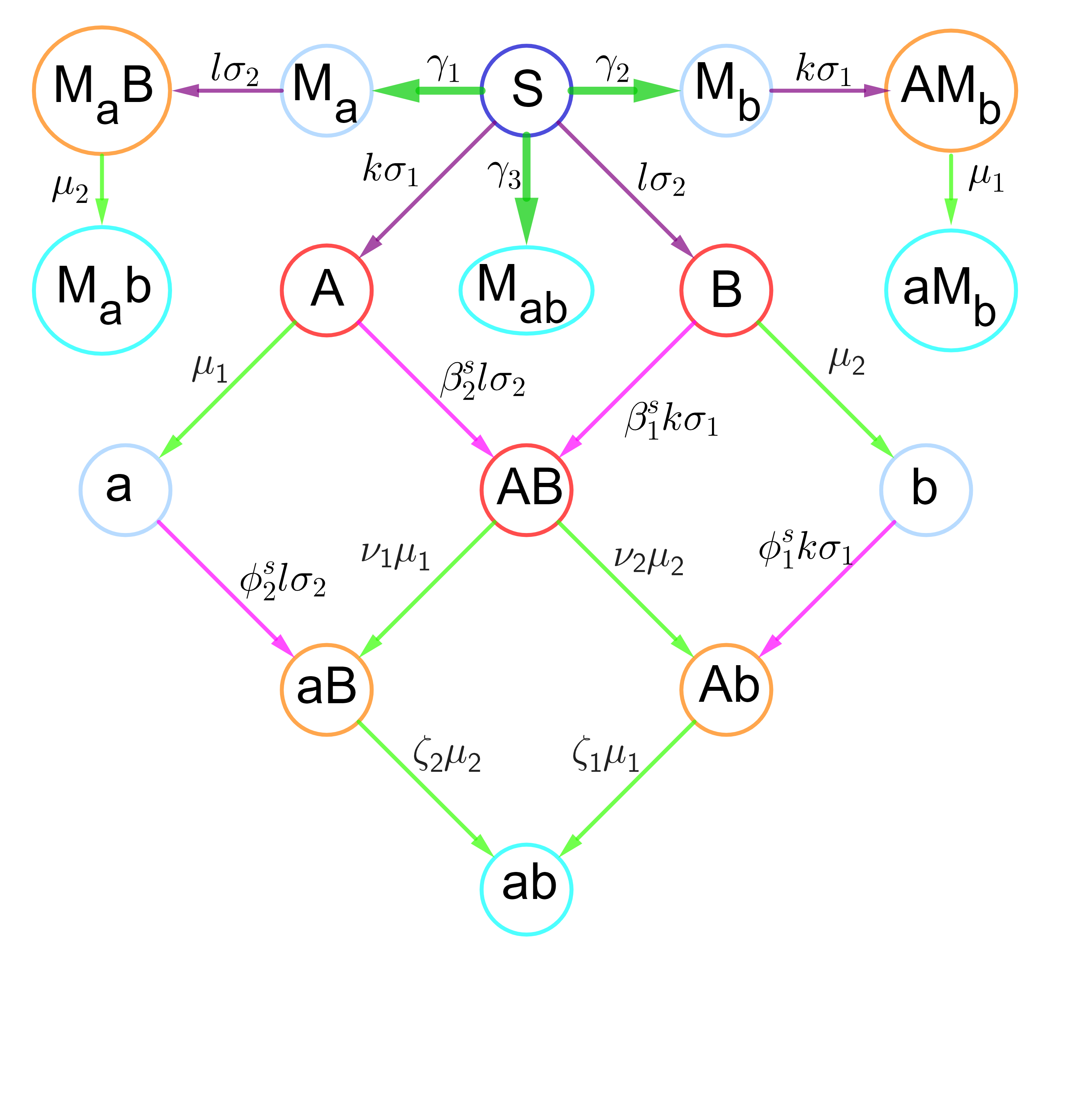}
			\caption{A schematic diagram for dynamics of two disease spreading on a population corresponding to a group with $k$ and $l$ degrees on networks on which the disease $A$ or $B$ spreads respectively. $M_{ab}$: vaccinated against both diseases. $M_a$ and $M_b$: vaccinated against one of the diseases. $AM_b$ and $aM_b$: vaccinated against $B$ and infected by or recovered from $A$. $M_aB$ and $M_ab$: vaccinated against $A$ and infected by or recovered from $B$. Other compartments are the same as figure~\ref{fig:2SIR}. The parameters are defined in Table~\ref{tab:ModelParam}. The dynamics obey equations~(\ref{eq:DBMF}). $\sigma_i$ and $\gamma_i$ are defined in equations~(\ref{eq:SigmaGamma}).}
			\label{fig:DBA2SIR}
		\end{center}
	\end{figure}

	Up until now, we have dealt with the case of symmetric spreads of co-infections on a well mixed population. In this section, we deviate from these assumptions and explore effects of measures on the dynamic.	We have seen, in the previous section, how measures with a rate proportional to the size of the infectious compartment can change the type of transitions in a symmetric, homogeneous case. If we forgo the symmetry and homogeneity, we cannot have analytical results as before, at least in a mean-field approach, however, we can make numerical experimentation to investigate if measures can show similar effects in such cases. Adopting a heterogeneous mean-field approximating approach, we can include network structures in the dynamics of spreading co-infections. In such methods, the population is divided into groups who have the same degree in the network. It is assumed that the status of all members in each group is always similar throughout the dynamic~\cite{Sanz2014}. Suppose there is a population in which diseases $A$ and $B$ are spreading. In the most general case, each disease can spread on its corresponding network. Hence, there are two networks on any of which the disease $A$ or $B$ spreads. Each block of the population is identified by a tuple $(k,l)$ corresponding to the degrees of that block on the two networks. The probability distribution of degrees on the networks is denoted by $P(k,l)$. A diagram illustrating the dynamic of the block $(k,l)$ status is depicted in figure~\ref{fig:DBA2SIR}.

	The probability of the block $(k,l)$ getting a new infection, for example $A$, changes at a rate which can be determined by multiplying the corresponding degree ($k$ for the disease $A$), the transmission probability ($\lambda_1$ for the disease $A$), the probability of being connected to a infectious node and the probability of being susceptible. The transmission probability might get reinforced (by parameters $\beta$ and $\phi$) if either of the nodes getting infected or the node that is the source of the infection has already been infected by the other disease. The recovery rate also might be different (using parameters $\nu$ and $\zeta$) for nodes that have got infected by both diseases. To determine the probability of a node connected to an infected node, we can multiply the probability of the connected node belonging to the block $(k,l)$ and the probability of that block being infected ($\theta$). The multiplication should be summed up over all $(k,l)$s. Also, let us define $\sigma_i$ and $\gamma_i$ as follows:
	
	\begin{equation}
		\def\arraystretch{2.0}
		\begin{array}{l} 	
			\sigma_1=\lambda_1(\theta^A_1+\beta^i_1\theta^{AB}_1+\phi^i_1\theta^{Ab}_1+\theta^{AM_b}_1), \\
			\sigma_2=\lambda_2(\theta^B_2+\beta^i_2\theta^{AB}_2+\phi^i_2\theta^{aB}_2+\theta^{M_aB}_2), \\
			\gamma_1=\eta_1(\theta^A_1+\theta^{AB}_1+\theta^{Ab}_1), \\
			\gamma_2=\eta_2(\theta^B_2+\theta^{AB}_2+\theta^{aB}_2), \\
			\gamma_3=\eta_3(\theta^A_1+\theta^{AB}_1+\theta^{Ab}_1+\theta^B_2+\theta^{AB}_2+\theta^{aB}_2). \\
		\end{array} 
		\label{eq:SigmaGamma}
	\end{equation}		
	
	The parameters used in the model are all defined in the Table~\ref{tab:ModelParam}. 
	Here, $\lambda_1$ and $\lambda_2$ are different from $\alpha$ (figure~\ref{fig:2SIR}) in a way that, for homogeneous symmetric cases, we have  $\alpha=\expval{k}\lambda_1=\expval{k}\lambda_2$. The variable $\theta$ represents the probability of being infectious at the neighborhood of a block (Table~\ref{tab:theta}). As you can see the rate at which the measures are applied in each block ($\gamma_i$) is proportional to the probability of being infectious at the neighborhood of the block [equations~(\ref{eq:SigmaGamma})]. Through the preceding steps and definitions, we can obtain the equations governing the dynamics as follows:

	\begin{equation}
		\def\arraystretch{2.0}
		\begin{array}{l} 	
		\dot{S}(k,l)=-(k\sigma_1+l\sigma_2)S(k,l)-(\gamma_1+\gamma_2+\gamma_3)S(k,l),
		\\
		\dot{A}(k,l)=k\sigma_1S(k,l)-\beta^s_2l\sigma_2A(k,l)-\mu_1A(k,l),
		\\		
		\dot{B}(k,l)=l\sigma_2S(k,l)-\beta^s_1k\sigma_1B(k,l)-\mu_2B(k,l),
		\\
		\dot{AB}(k,l)=\beta^s_1k\sigma_1B(k,l)+\beta^s_2l\sigma_2A(k,l) \\
		\hspace{1.75cm} -(\nu_1\mu_1+\nu_2\mu_2)AB(k,l),
		\\		
		\dot{a}(k,l)=\mu_1A(k,l)-\phi^s_2l\sigma_2a(k,l),
		\\
		\dot{b}(k,l)=\mu_2B(k,l)-\phi^s_1k\sigma_1b(k,l),
		\\		
		\dot{aB}(k,l)=\phi^s_2l\sigma_2a(k,l)+\nu_1\mu_1AB(k,l)-\zeta_2\mu_2aB(k,l),
		\\
		\dot{Ab}(k,l)=\phi^s_1k\sigma_1b(k,l)+\nu_2\mu_2AB(k,l)-\zeta_1\mu_1Ab(k,l),
		\\
		\dot{M_a}(k,l)=\gamma_1S(k,l)-l\sigma_2M_a(k,l),
		\\
		\dot{M_b}(k,l)=\gamma_2S(k,l)-k\sigma_1M_b(k,l),
		\\
		\dot{M_aB}(k,l)=l\sigma_2M_a(k,l)-\mu_2M_aB(k,l),
		\\
		\dot{AM_b}(k,l)=k\sigma_1M_b(k,l)-\mu_1AM_b(k,l),
		\\
		\dot{M_{ab}}(k,l)=\gamma_3S(k,l),
		\\
		\dot{M_{a}b}(k,l)=\mu_2M_aB(k,l),
		\\
		\dot{aM_{b}}(k,l)=\mu_1AM_b(k,l),
		\\
		\dot{ab}(k,l)=\zeta_1\mu_1Ab(k,l)+\zeta_2\mu_2aB(k,l).\\				
		
		\end{array} 
		\label{eq:DBMF}
	\end{equation}

\begin{figure*}[htbp]
		\begin{center}
			\includegraphics[width=6.5in]{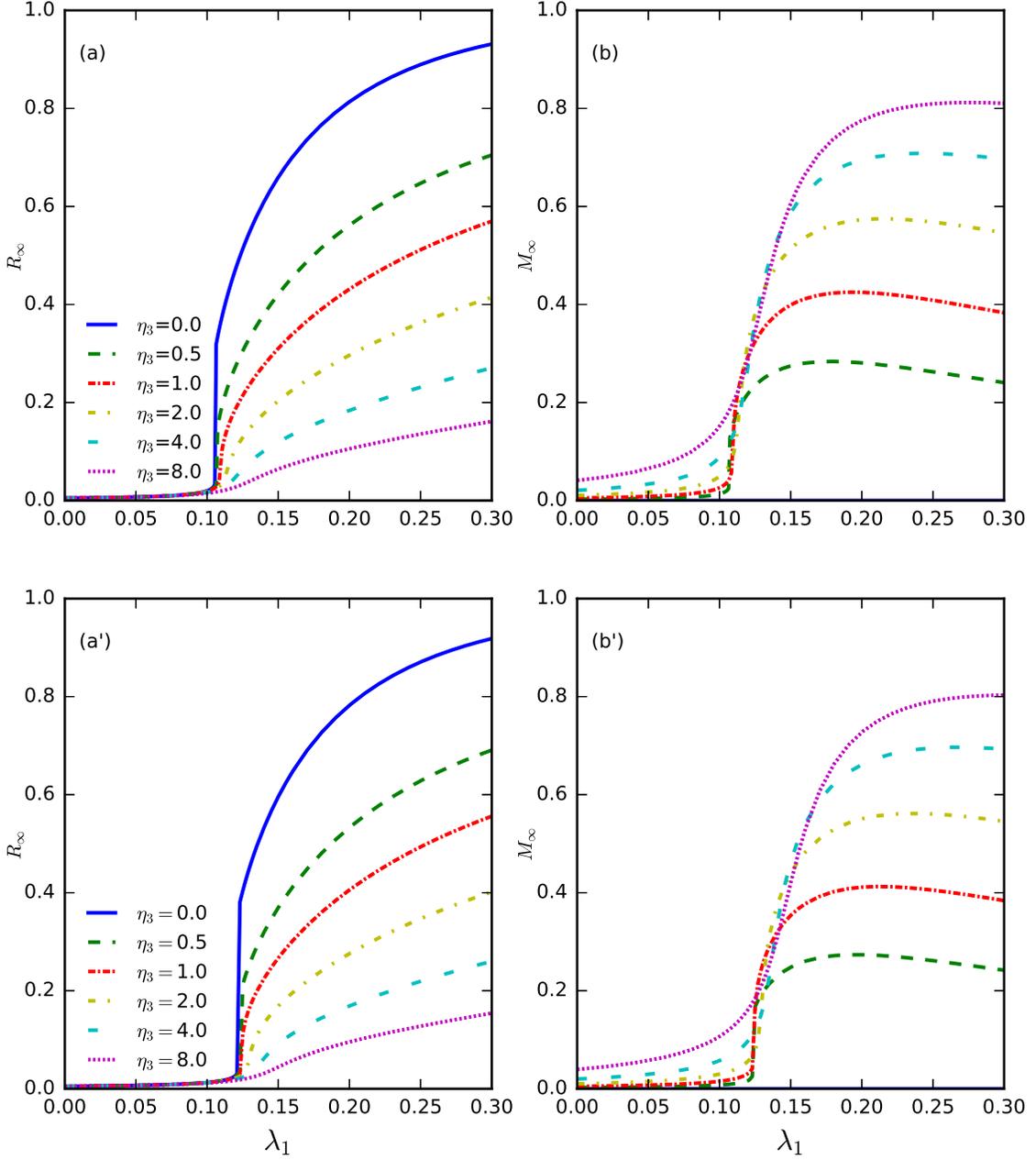}
			\caption{The effect of measures with a rate proportional to the probability of the existence of infections among neighbors on phase transitions in symmetric heterogeneous cases of two coupled SIR dynamics. The recovered and the vaccinated at the end of the dynamic, $R_\infty$ and $M_\infty$ defined by equation~(\ref{eq:RinfMinf}), for the spread of diseases on two networks, $N=5000, \expval{k}=\expval{l}=5$, are depicted against the infectiousness of the disease $A$ in the absence of the disease $B$ ($\lambda_1$) for different measure rates $\eta_3$. Model parameters are $\lambda_2=\lambda_1, \beta^s_1=\beta^s_2=45, \phi^s_1=\phi^s_2=30$ [equations~(\ref{eq:DBMF})]. Other parameters regarding infection rates and all recovery parameters are set equal to 1. The population is vaccinated against both diseases. The initial seeds are distributed homogeneously in a way that $\expval{S_0}=0.995, \expval{A_0}=0.0025$ and $\expval{B_0}=0.0025$. (a) and (a$'$) $R_\infty$, (b) and (b$'$) $M_\infty$. (a) and (b) Scale Free networks [$P(k)\sim k^{-2.1}$], (a$'$) and (b$'$) Erdos-Renyi networks.}
			\label{fig:DBA_ER_SC_S5}
		\end{center}
	\end{figure*}

\begin{figure*}[htbp]
		\begin{center}
			\includegraphics[width=6.5in]{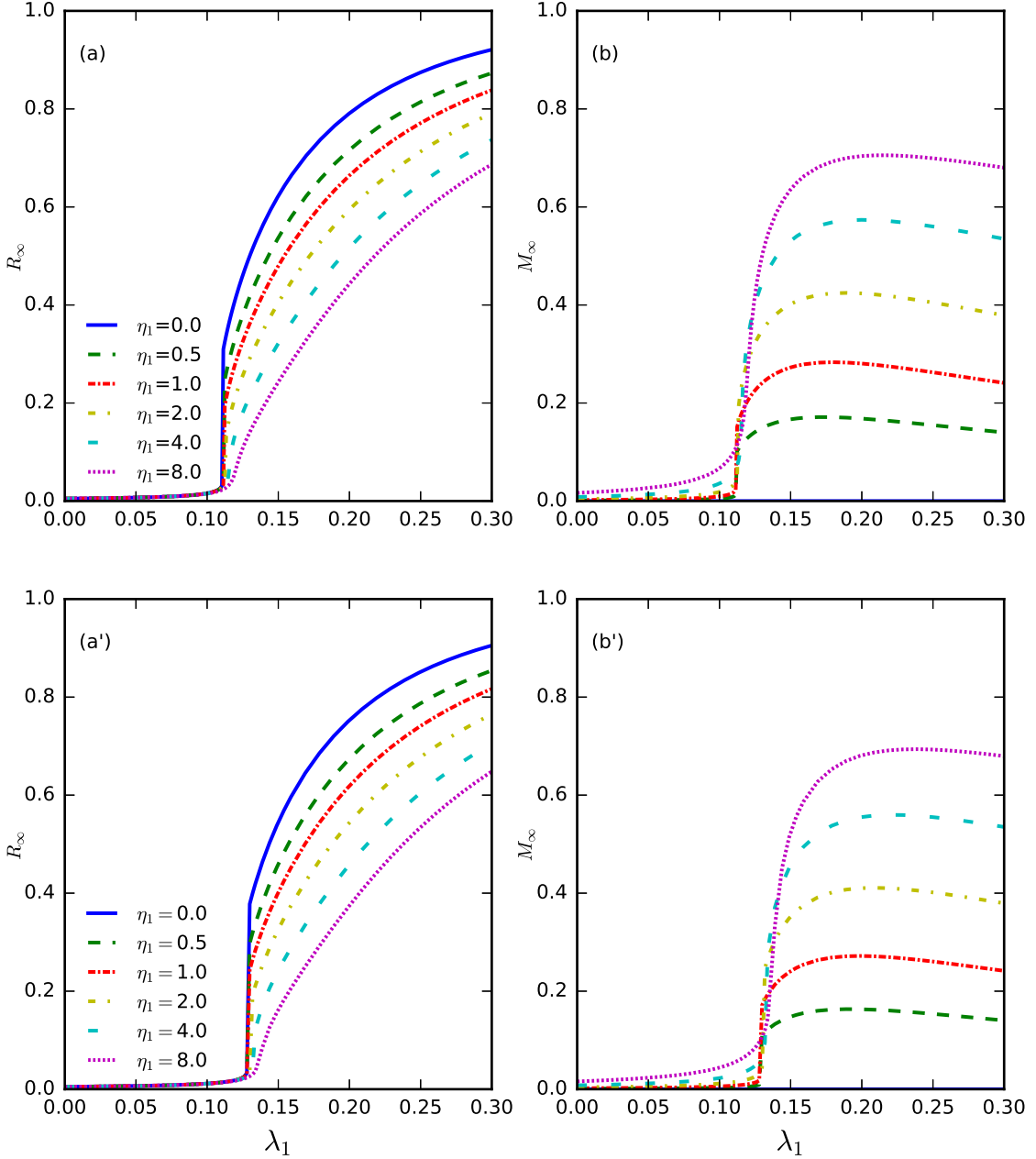}
			\caption{Similar to figure~\ref{fig:DBA_ER_SC_S5} for asymmetric heterogeneous cases. Diagrams of $R_\infty$ and $M_\infty$ are depicted against $\lambda_1$ for different values of $\eta_1$. Model parameters are $\lambda_2=0.9\lambda_1, \beta^s_1=30, \beta^s_2=60, \phi^s_1=21$ and $\phi^s_2=48$. Other parameters regarding infection rates and all recovery parameters are set equal to 1. The population is vaccinated against the disease $A$. The initial seeds are distributed homogeneously in a way that $\expval{S_0}=0.995, \expval{A_0}=0.002$ and $\expval{B_0}=0.003$. (a) and (a$'$) $R_\infty$, (b) and (b$'$) $M_\infty$. (a) and (b) Scale Free networks [$P(k)\sim k^{-2.1}$], (a$'$) and (b$'$) Erdos-Renyi networks.}
			\label{fig:DBA_ER_SC_U5}
		\end{center}
	\end{figure*}

	 Using the framework explained above and through numerical experiments, we can examine if an asymmetry or heterogeneity might dramatically change the effects of measures observed in the symmetric case. In results illustrated in figure~\ref{fig:DBA_ER_SC_S5} and~\ref{fig:DBA_ER_SC_U5}, the average steady state size of the recovered ($R_\infty$) and the vaccinated ($M_\infty$) are plotted against different values of the infectiousness of the disease $A$ in the absence of the disease $B$ ($\lambda_1$). $M_\infty$ and $R_\infty$ are defined as follows:
	
	\begin{equation}
		\def\arraystretch{1.5}
		\begin{array}{l} 	
		M_\infty=\sum\limits_{k,l}P(k,l)\left(M_{ab}+M_a+M_ab+M_b+aM_b\right)(k,l), \\
		R_\infty=\sum\limits_{k,l}P(k,l)\left(a+b+ab+M_ab+aM_b\right)(k,l).
		\end{array} 
		\label{eq:RinfMinf}
	\end{equation}	

	\begin{table*}[ht]
		\caption{Definition of model parameters.}
		
		\begin{tabularx}{\textwidth}{@{\hspace{18pt}} l @{\hspace{18pt}} X }  
			\hline \hline
			Parameter & Definition
			\\ \hline
			$\lambda_1$ & Disease A infectiousness in the absence of B.   \\
			$\lambda_2$ & Disease B infectiousness in the absence of A.  \\
			$\mu_1$ & Disease A recovery rate in the absence of B.  \\
			$\mu_2$ & Disease B recovery rate in the absence of A.  \\
			$\beta^{s}_1$ & Increase of disease A infectiousness because the susceptible individual is already infected by disease B.  \\
			$\beta^{s}_2$ & Increase of disease B infectiousness because the susceptible individual is already infected by disease A.  \\
			$\phi^{s}_1$ & Increase of disease A infectiousness because the susceptible individual has been recovered from disease B.  \\
			$\phi^{s}_2$ & Increase of disease B infectiousness because the susceptible individual has been recovered from disease A.  \\
			$\beta^{i}_1$ & Increase of disease A infectiousness because the infectious individual is also infected by disease B.  \\
			$\beta^{i}_2$ & Increase of disease B infectiousness because the infectious individual is also infected by disease A.  \\
			$\phi^{i}_1$ & Increase of disease A infectiousness because the infectious individual has been recovered from disease B.  \\
			$\phi^{i}_2$ & Increase of disease B infectiousness because the infectious individual has been recovered from disease A.  \\
			$\nu_1$ & Change of disease A recovery rate because the individual is also infected by disease B.  \\
			$\nu_2$ & Change of disease B recovery rate because the individual is also infected by disease A.  \\
			$\zeta_1$ & Change of disease A recovery rate because the individual has also been recovered from disease B.   \\
			$\zeta_2$ & Change of disease B recovery rate because the individual has also been recovered from disease A.  \\
			$\eta_1$ & Rate of immunization against disease A.   \\
			$\eta_2$ & Rate of immunization against disease B.   \\
			$\eta_3$ & Rate of immunization against both diseases.   \\
			\hline\hline
		\end{tabularx}
		
		\label{tab:ModelParam}
	\end{table*}

	\begin{table*}[ht]
		\caption{Definition of the parameters $\theta$.}
		
		\begin{tabularx}{\textwidth}{@{\hspace{18pt}} l @{\hspace{18pt}} X }  
			\hline \hline \\
			$\theta^A_1=\frac{1}{\expval{k}}{\sum\limits_{k,l}(kP(k,l)A(k,l))}$ & Probability that a node on network 1 is connected to an A node.   \\
			$\theta^{AB}_1=\frac{1}{\expval{k}}{\sum\limits_{k,l}(kP(k,l)AB(k,l))}$ & Probability that a node on network 1 is connected to an $AB$ node.  \\
			$\theta^{Ab}_1=\frac{1}{\expval{k}}{\sum\limits_{k,l}(kP(k,l)Ab(k,l))}$ & Probability that a node on network 1 is connected to an $Ab$ node.  \\
			$\theta^{AM_b}_1=\frac{1}{\expval{k}}{\sum\limits_{k,l}(kP(k,l)AM_b(k,l))}$ & Probability that a node on network 1 is connected to an $AM_b$ node.  \\			
			$\theta^{B}_2=\frac{1}{\expval{l}}{\sum\limits_{k,l}(lP(k,l)B(k,l))}$ & Probability that a node on network 2 is connected to a $B$ node.  \\			
			$\theta^{AB}_2=\frac{1}{\expval{l}}{\sum\limits_{k,l}(lP(k,l)AB(k,l))}{\expval{l}}$ & Probability that a node on network 2 is connected to an $AB$ node.  \\
			$\theta^{aB}_2=\frac{1}{\expval{l}}{\sum\limits_{k,l}(lP(k,l)aB(k,l))}$ & Probability that a node on network 2 is connected to an $aB$ node.  \\
			$\theta^{M_aB}_2=\frac{1}{\expval{l}}{\sum\limits_{k,l}(lP(k,l)M_aB(k,l))}$ & Probability that a node on network 2 is connected to a $M_aB$ node.  \\
			\hline\hline
		\end{tabularx}
		\label{tab:theta}
	\end{table*}

\begin{figure*}[htbp]
	\begin{center}
		\includegraphics[width=6.5in]{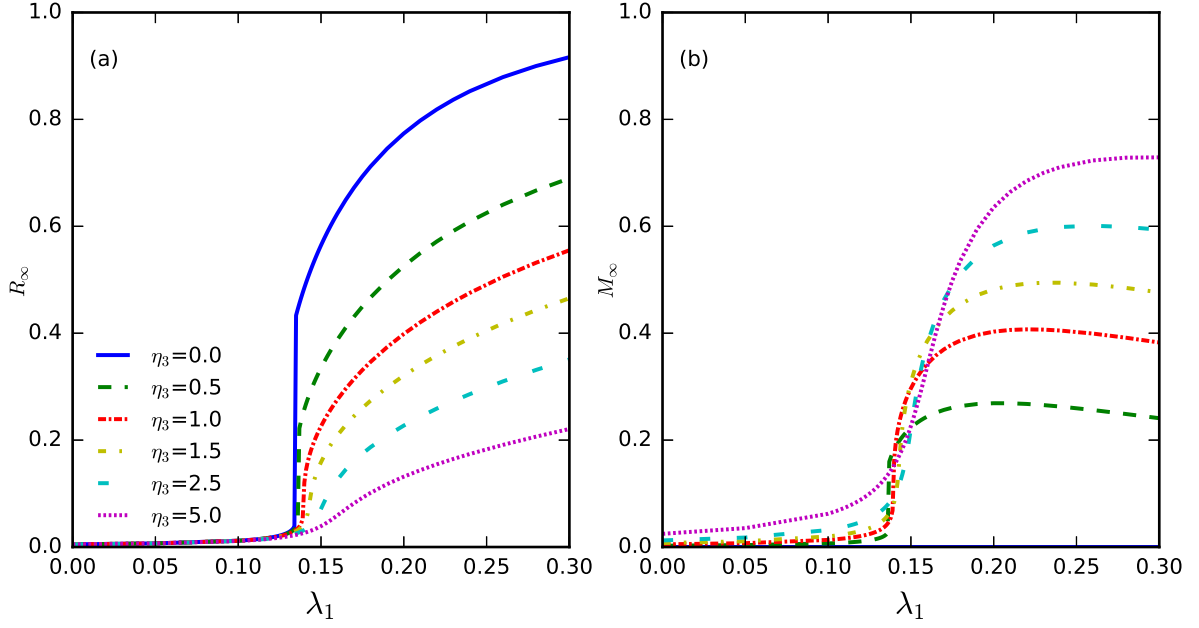}
		\caption{Phase transitions of a symmetric spread of two cooperative diseases in a heterogeneous population with the same parameters as the case of the well-mixed population presented in figure~\ref{fig:RVvsAlphaEtaX} for a comparison between the two. The recovered and the vaccinated at the end of the dynamic, $R_\infty$ and $M_\infty$ defined by equation~(\ref{eq:RinfMinf}), for the spread of diseases on two Erdos-Renyi networks , $N=5000, \expval{k}=\expval{l}=5$, are depicted against the infectiousness of the disease $A$ in the absence of the disease $B$ ($\lambda_1$) for different measure rates $\eta_3$. Model parameters are $\lambda_2=\lambda_1, \beta^s_1=\beta^s_2=15, \phi^s_1=\phi^s_2=15$ [equations~(\ref{eq:DBMF})]. Other parameters regarding infection rates and all recovery parameters are set equal to 1. The population is vaccinated against both diseases. The initial seeds are distributed homogeneously in a way that $\expval{S_0}=0.995, \expval{A_0}=0.0025$ and $\expval{B_0}=0.0025$. (a) $R_\infty$, (b) $M_\infty$.}
			\label{fig:HomoComprison}
	\end{center}
\end{figure*}

\begin{figure*}[htbp]
	\begin{center}
		\includegraphics[width=6.5in]{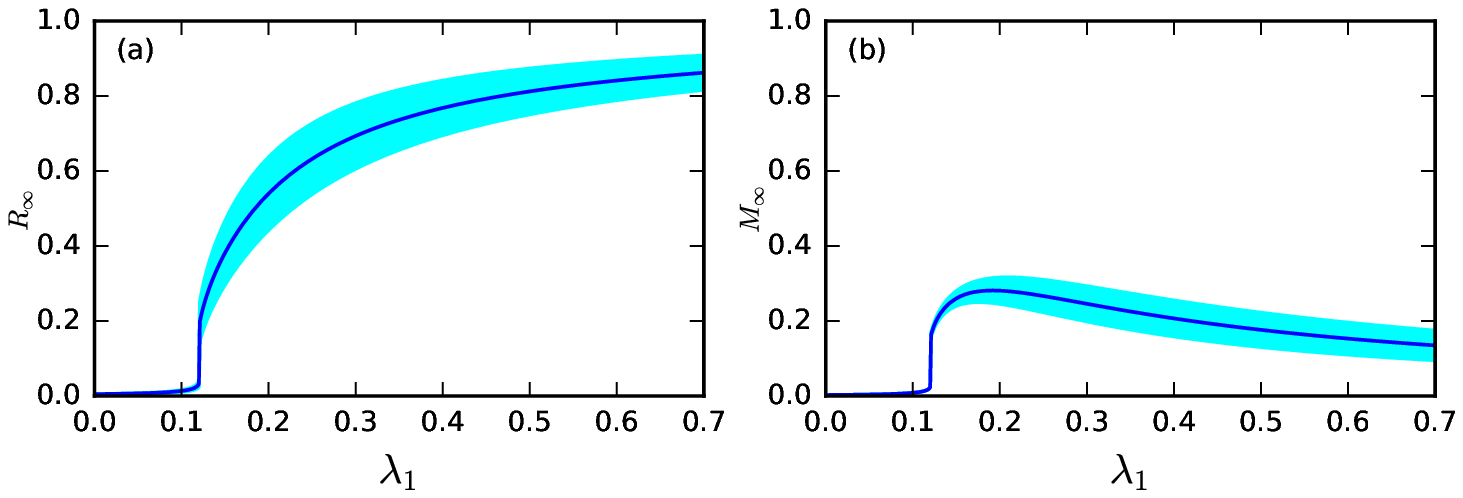}
		\caption{Illustration of the standard deviation of the recovered and the vaccinated in one of the diagrams of figure~\ref{fig:DBA_ER_SC_S5}(a$'$) and~\ref{fig:DBA_ER_SC_S5}(b$'$), $\eta_3=0.5$.}
			\label{fig:DBA_SD}
	\end{center}
\end{figure*}

	Two networks with $N=5000$ nodes and mean degrees of $\expval{k}=\expval{l}=5$ are chosen as networks on which the diseases are spreading. In figure~\ref{fig:DBA_ER_SC_S5}, the parameters are symmetric and the vaccination is applied against both diseases. Meanwhile, in figure~\ref{fig:DBA_ER_SC_U5}, the infectivity parameters are not symmetric and the population is vaccinated only against disease $A$. Initial seeds are spread homogeneously among all blocks. The left and right panels correspond to the average of recovered compartments $R_\infty$ and the average of measures compartments $M_\infty$, respectively. Moreover, the top (a),(b) and bottom (a$'$),(b$'$) panels correspond to scale-free networks [$P(k)\sim k^{-2.1}$] and Erdos-Renyi networks, respectively.

	As shown in the results, the change of transition types due to the measures are similar to the symmetric homogeneous cases. Even applying measures only to one disease can turn the discontinuous transition into a continuous one if their rates ($\eta$) are high enough [figure~\ref{fig:DBA_ER_SC_U5}]. The results show that applying the measures symmetrically evidently leads to a smaller $R_{\infty}$ and more vaccination for the same $\lambda_1$ (for instance, compare the diagram of $R_\infty$ and $M_\infty$ corresponding to $\eta_3=1$ and $\eta_1=1$ in figures~\ref{fig:DBA_ER_SC_S5} and~\ref{fig:DBA_ER_SC_U5}, respectively). Comparing the top and bottom panels, corresponding to scale-free and Erdos-Renyi networks, respectively, illustrates that the threshold is slightly smaller for scale-free networks. Numerical experimentation performed with a heterogeneous spread of initial seeds, not shown in the results, demonstrated similar characteristics for transitions. \\
	We have also looked at a case whose parameters are the same as those used to determine the results in figure \ref{fig:RVvsAlphaEtaX} so that we can compare the case of a well-mixed population with the case of a heterogeneous population. We have set $\lambda_2=\lambda_1$ and $\beta^s_1=\beta^s_2=\phi^s_1=\phi^s_2=15$. Other parameters regarding infection rates and all recovery parameters are set equal to 1. The population is vaccinated against both diseases. The initial seeds are distributed homogeneously in a way that $\expval{S_0}=0.995$ and $\expval{A_0}=\expval{B_0}=0.0025$. Two Erdos-Renyi networks with $N=5000$ nodes and mean degrees of $\expval{k}=\expval{l}=5$ are selected as networks on which the diseases are spreading. The results are presented in figure~\ref{fig:HomoComprison}. The average of recovered compartments $R_\infty$ and the average of measures compartments $M_\infty$ have been depicted against $\lambda_1$ for different values of $\eta_3$. To compare figure~\ref{fig:RVvsAlphaEtaX} with figure~\ref{fig:HomoComprison}, we should note the difference between $\alpha$ and $\lambda_1$ and between $\eta$ and $\eta_3$. According to the definitions, $\lambda_1$ actually corresponds to $\frac{\alpha}{\expval{k}}$. Furthermore, if we compare the relation $\gamma=\eta X$ of section \ref{sec:analytical} with the definition of $\eta_3$ in equation~\ref{eq:SigmaGamma}, we see that $\eta_3$ corresponds to $\frac{\eta}{2}$. Comparing figure~\ref{fig:RVvsAlphaEtaX} with figure~\ref{fig:HomoComprison}, we can see that the heterogeneity decreases the phase-transition threshold. For instance, the threshold for the case of no measure has been decreased from $\lambda_1=\frac{\alpha}{\expval{k}}=0.158$ for a well-mixed population (figure~\ref{fig:RVvsAlphaEtaX}) to $\lambda_1=0.134$ in a heterogeneous one (figure~\ref{fig:HomoComprison}). Moreover, the minimum vaccination rate which can turn the discontinuous transition into a continuous one has been also decreased by the heterogeneity. In figure~\ref{fig:AnalRvsAlpha}, we have seen that such a vaccination rate for the case of a well-mixed population is $\eta=3.26$. In figure~\ref{fig:HomoComprison}, on the other hand, the discontinuous transition turns into a continuous one for $\eta_3$ between 1 and 1.5, which is lower than the corresponding $\eta_3=\frac{\eta}{2}$. \\
	It should be mentioned that every block $(k,l)$ undergoes its specific time evolution and has its own steady-state size of the recovered and the vaccinated which can differ from the average values  $R_\infty$ and $M_\infty$. The standard deviation of the values has been determined. As an instance, the standard deviation is illustrated in figure~\ref{fig:DBA_SD}, which shows that the range of values corresponding to blocks is small for small $\lambda_1$ and is maximum around the transition $\lambda_1$.

\section{Concluding remarks}
\label{sec:conclusion}
	
	The effect of control measures that could remove individuals from the susceptible, like vaccination, was introduced into the mean-field equations governing the SIR dynamics of cooperative co-infections. Through numerical experiments, the competition between measures and diseases to immunize the population was explored. Results showed that the presence of measures increases the threshold of epidemics. Moreover, when the rate of measures exceeds a certain value, the discontinuous transitions observed in cooperative co-infections could turn into continuous ones. Assuming that the authority's policies and people's attitudes regarding control measures could make the measure rate dictated by the size of the infectious population, we scrutinized a dynamic in which the measure rate is proportional to the size of the infectious compartment. We showed that vaccination with such a variable rate can change the type of transitions too. Moreover, we managed to solve the governing equations analytically. Analytical solutions show how the saddle-node-like bifurcation corresponding to a discontinuous transition in the dynamic, observed in a previous study~\cite{Zarei2019}, could be eliminated when the measure rate reaches a threshold. Furthermore, through analytical results, systems of equations are obtained whose solutions give the threshold for transitions and the measure rate at which the transition type changes. Two differences show up between the effects of fixed-rate and variable-rate measures. First, with fixed-rate measures, the threshold of the phase transition increases much more than measures with variable rates. Second, with fixed-rate measures, the size of the measure compartment at the end of the dynamic peaks at the vicinity of the threshold, while it keeps rising after the threshold for variable-rate measures. The effects of measures were also incorporated in a framework which includes the structure of networks in the dynamics (using a heterogeneous mean-field approximation). Numerical experiments showed that even vaccinating the population against one of diseases could change the type of transitions from discontinuous to continuous. We also compared phase transitions of the symmetric spread of two cooperative diseases in a well-mixed and in a heterogeneous one. The results show that while the heterogeneity has the adverse effect of decreasing the transition threshold, the minimum vaccination rate at which the discontinuous transition becomes continuous decreases too.\\
	Our study was limited to dynamics in which the measures are applied during the spread of the diseases. There are some aspects that may need further examinations. In all of our results, we assumed that both diseases start spreading at the same time. We adopted the mean-field approach. Simulating the dynamics on different networks could provide more insight into the interplay of propagating diseases and the vaccination process. We applied the measures homogeneously among the population. Investigating the effect of different vaccination strategies can be a subject of interest. Here, the people entering the measures compartment could never go back to be susceptible. Extending the model to eliminate this limitation, one can probe effects of imperfect vaccination or quarantine. Diving into applications of the model to real cases can be interesting, which has not been explored in our study. In that regard, the time scale of the diseases and the range of infection, recovery and vaccination rates, specially relative to each other, should be investigated thoroughly. In this study we have focused on the epidemic thresholds and the type of transitions, however, we have also observed that vaccination postpones the apogee of the infectious population while decreasing the peak value, which is favorable in controlling and suppressing an epidemic and can establish a criterion for the optimum rate of vaccination. Further studies can address this subject.
	
\begin{acknowledgments}
	F. Gh. acknowledges partial support by Deutsche Forschungsgemeinschaft (DFG) under the grant project: 345463468 (idonate). 
\end{acknowledgments}
	
\appendix*
\section{Supplementary Figures}
\begin{figure*}[htbp]
	\begin{center}
	\includegraphics[width=6.5in]{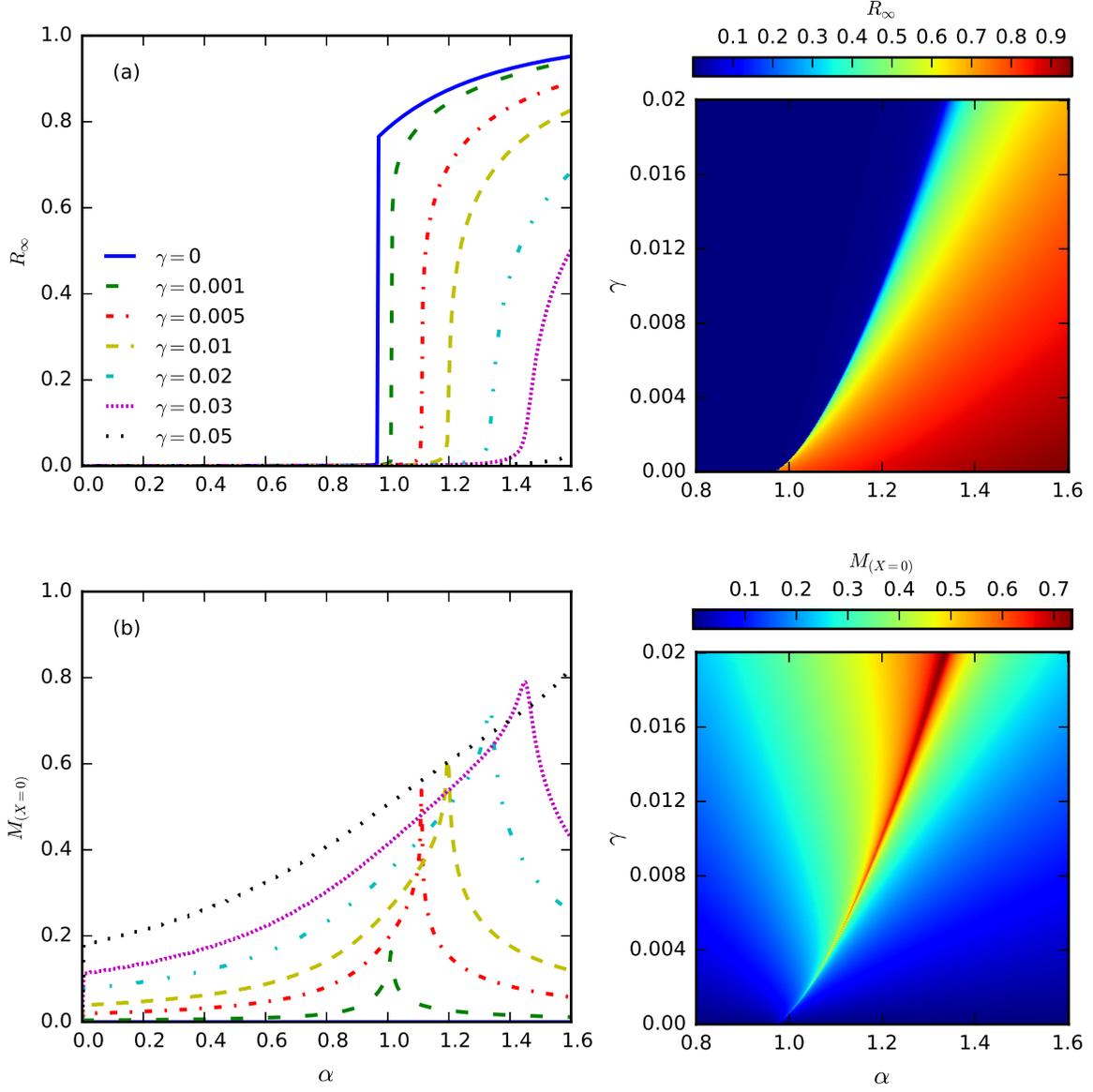}
		\caption{Similar to figure~\ref{fig:RVvsAlpha} for $\epsilon=0.0001$.}
		\label{fig:RVvsAlpha1}
	\end{center}
\end{figure*}

\begin{figure*}[h]
	\begin{center}
	\includegraphics[width=6.5in]{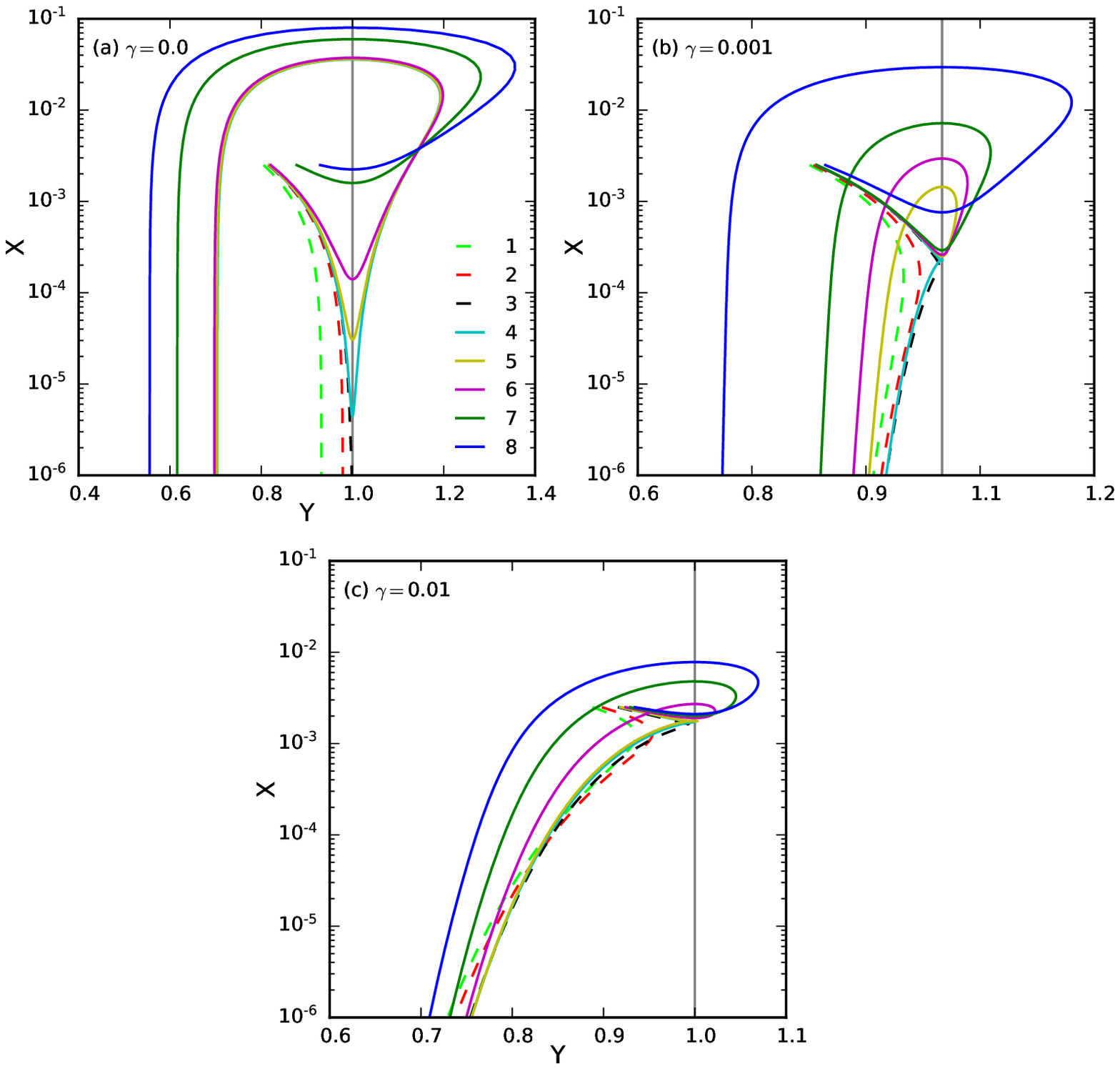}
		\caption{Near-threshold trajectories of $X(t)$ are depicted against $Y(t)$ for different values of $\gamma$. The model parameters and initial conditions are $\beta=15\alpha$, $X_0=P_0=\frac{\epsilon}{2}$, $S_0=1-\epsilon$ and $\epsilon=0.005$ (equations~(\ref{eq:SPXV})). Numbers 1-8 in the legend correspond to values of $\alpha$ as follows. (a) $\alpha \in \{0.78, 0.79, 0.7912, 0.7913, 0.792, 0.795, 0.85, 0.90\}$, (b) $\alpha \in \{0.80, 0.805, 0.8081, 0.8083, 0.8085, 0.8086, 0.809, 0.82\}$, (c) $\alpha \in \{0.86, 0.87, 0.887, 0.8892, 0.89, 0.895, 0.90, 0.905\}$ . Dashed and solid lines correspond to cases with $\alpha$ below and above the threshold respectively.}
		\label{fig:XvsY}
	\end{center}
\end{figure*}

\begin{figure*}[h]
	\begin{center}
	\includegraphics[width=6.5in]{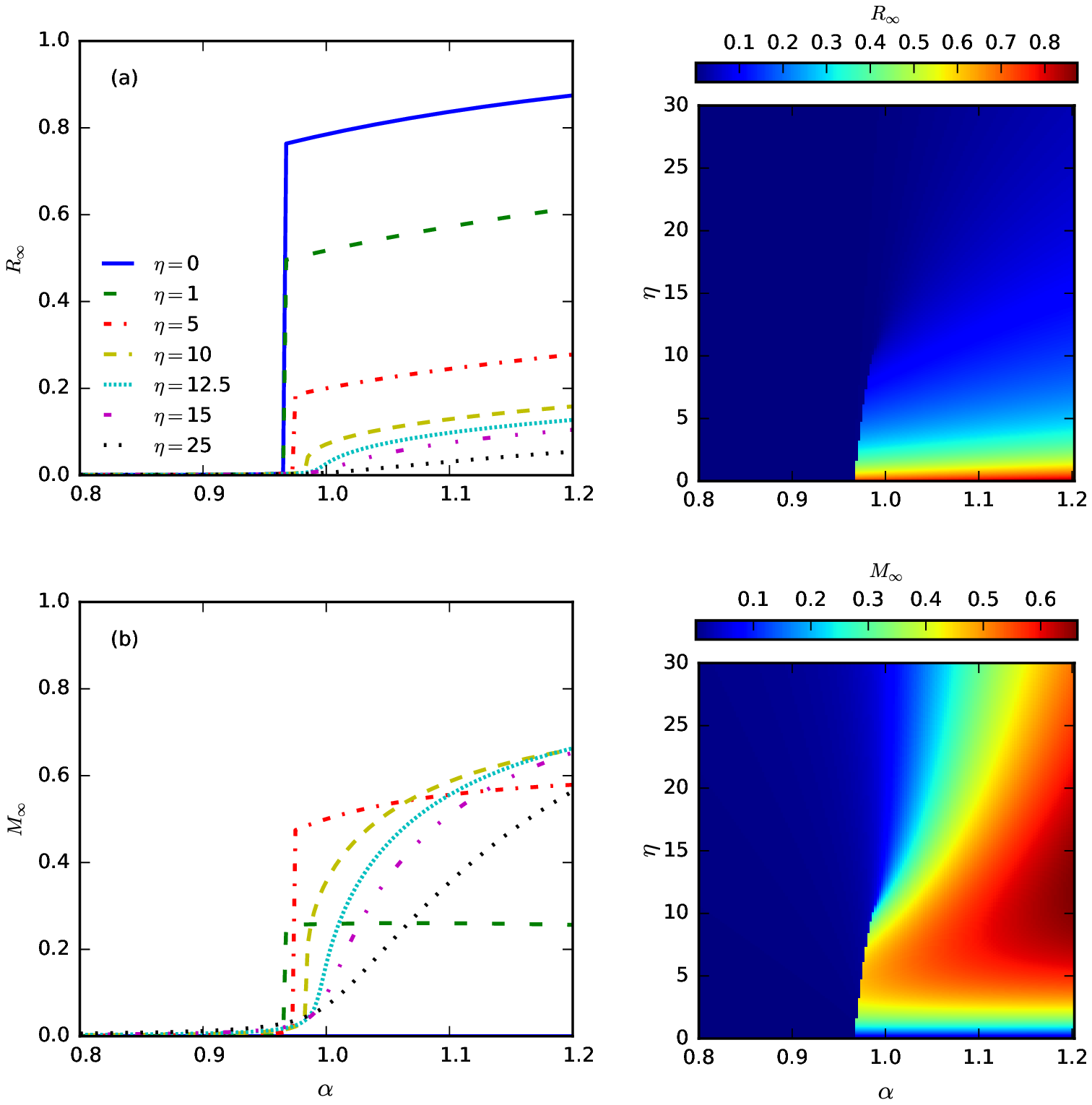}
		\caption{Similar to figure~\ref{fig:RVvsAlphaEtaX} for $\epsilon=0.0001$.}
		\label{fig:RVvsAlphaEtaX1}
	\end{center}
\end{figure*}

\begin{figure*}[t]
	\begin{center}
	\includegraphics[width=6.5in]{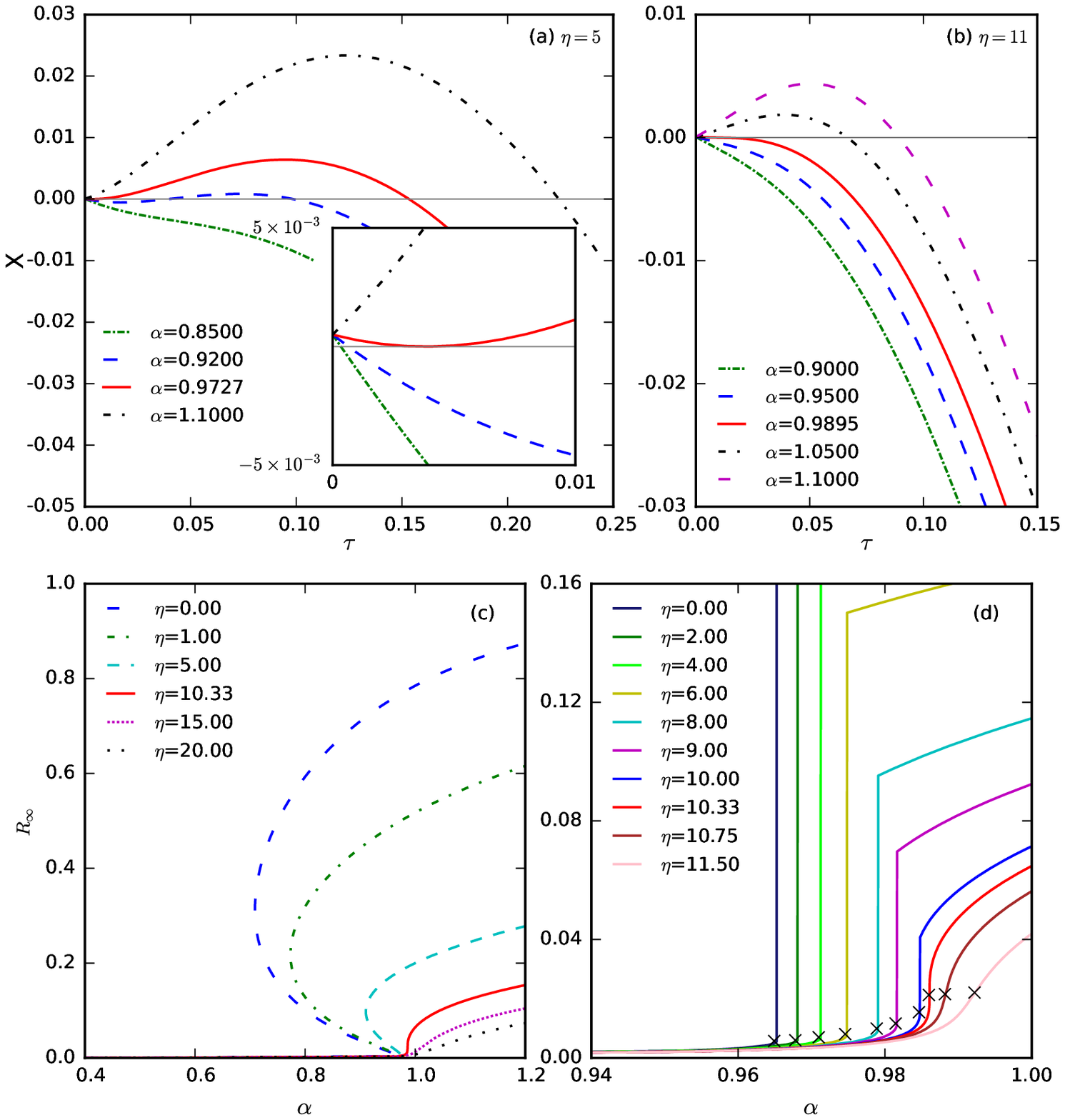}
		\caption{Results determined from analytical solution to equations~(\ref{eq:SPXVEtaX}), (a) and (b) similar to figure~\ref{fig:XTauBifur} for $\epsilon=0.0001$, (c) similar to figure~\ref{fig:AnalRvsAlpha} for $\epsilon=0.0001$, (d) similar to figure~\ref{fig:Tricirtical} for $\epsilon=0.0001$.}
		\label{fig:XTauBifur1}
	\end{center}
\end{figure*}

In sections \ref{sec:2SIR_Vaccination} and \ref{sec:analytical}, we have figures illustrating solutions to the governing equations of the dynamics for initial conditions $X_0=P_0=\frac{\epsilon}{2}$, $S_0=1-\epsilon$, and $\epsilon=0.005$. Here, we present solutions for a lower level of initial infections by setting $\epsilon=0.0001$ to supplement the results (figures ~\ref{fig:RVvsAlpha1}, ~\ref{fig:RVvsAlphaEtaX1}, and ~\ref{fig:XTauBifur1}). We have also included near-threshold trajectories of the dynamics governed by equations~(\ref{eq:SPXV}) in $Y-X$ plane (figure~\ref{fig:XvsY}).

\clearpage
\bibliographystyle{ieeetr}
	\bibliography{Effects_of_measures_on_phase_transitions_in_two_cooperative_Susceptible_Infectious_Recovered_SIR_dynamics}

\end{document}